# Hyperelasticity of Blood Clots: Bridging the Gap between Microscopic and Continuum Scales

Nicholas Filla[1], Beikang Gu[1], Jixin Hou[1], Kenan Song[1], He Li[2], Ning Liu[3] and Xianqiao Wang[1]

[1]School of Environmental, Civil, Agricultural and Mechanical Engineering, College of Engineering, University of Georgia, Athens, GA, 30602

[2]School of Chemical, Materials and Biomedical Engineering, College of Engineering, University of Georgia, Athens, GA, 30602

[3]School of Aerospace Engineering and Applied Mechanics, Tongji University, Shanghai 200092, P. R. China

Corresponding authors: xqwang@uga.edu

**Abstract** The biomechanical properties of blood clots, which are dictated by their compositions and micro-structures, play a critical role in determining their fates, i.e., occlusion, persistency, or embolization in the human circulatory system. While numerous constitutive models have emerged to describe the biomechanics of blood clots, the majority of these models have primarily focused on the macroscopic deformation of the clots and the resultant strain-stress correlations without depicting the microscopic contributions from their structural components, such as fibrin fibers, fibrin network and red blood cells. This work addresses the gap in current scientific understanding by quantifying how changes in the microstructure of blood clots affect its mechanical responses under different external stresses. We leverage our previous published work to develop a hyperelastic potential model for blood clots, which incorporates six distinct strain-energy components to describe the alignment of fibers, the entropic and enthalpic stretching of fibrin fibers, the buckling of these fibers, clot densification, and clot jamming. These strain-energy components are represented by a combination of simple harmonic oscillators, one-sided harmonic potentials, and a Gaussian potential. The proposed model, which is $C^0$, $C^1$, and $C^2$ continuous with a total of 13 parameters, has been validated against three data sets: 1) fibrin clot in tension, 2) blood clot in compression, and 3) blood clots in shear, demonstrating its robustness. Subsequent simulations of a microscopic blood clot model are performed to uncover mechanistic correlations for a majority of the hyperelastic potential's stiffness/strain parameters. Our results show that only one proposed term concerning fiber buckling needs further refinement, while the remaining five strain-energy terms appear to describe precisely what they were intended to. In summary, the proposed model provides insight into the behavior of thromboembolisms and assistance in computer-aided design of surgical tools and interventions such as thrombectomy.

# 1. Introduction

Blood clots, which play a crucial role in maintaining hemostasis and causing various vascular diseases, have intricate structures characterized by a porous mesh of branching fibrin fibers, entrapped red blood cells (RBCs), and platelets bound by fibrin. The mechanical properties of these clots, influenced by the arrangement and composition of their structural components, are of paramount importance for maintaining their physiological functions. Under normal conditions, the clots dissolve after the vessel damage is healed to restore the blood perfusion. However, under pathological conditions, the formation of obstructive clots, termed thrombi, gives rise to a spectrum of pathologies. When these thrombi persist without dissolution, they pose significant threats by potentially impeding blood flow, leading to conditions such as ischemic heart disease and ischemic stroke, which stand as the top two causes of global mortality [1, 2]. The exploration of the dynamics and interplay of blood clots stands as a critical facet of hematology research. Advancing our understanding of the role of fibrin fibers in blood clotting is foundational to improve treatments and therapies for disorders associated with abnormal blood clotting.

The mechanical properties of blood clots are essential not only in determining their *in vivo* functions but also in influencing their responses to medical procedures. In particular, the procedure of intra-arterial mechanical thrombectomy, which involves the removal of obstructions using stent retrievers or aspiration catheters, is significantly influenced by the mechanical characteristics of thrombi [3, 4]. This underscores the importance of comprehending clot composition, structure, and the resulting mechanical behavior. Despite its significance, the precise details of clot elasticity are currently a subject of debate, and the development of material constitutive models for clot elasticity has proven to be a challenging endeavor.

Prior studies of clot mechanics based on classical theories such as Neo-Hookean, Ogden, Yeoh, and Hyperfoam models have shown that these frameworks fall short in accurately capturing the complex mechanical behavior of fibrin networks and blood clots [5]. Unlike these conventional hyperelastic models, theoretical and numerical models for fibrin networks and blood clots require incorporating elements reflective of the force-extension behavior exhibited by fibrin fibers reported from different experimental studies [5-10]. Notably, there is an absence of precise models for the bilinear force-extension characteristics of individual fibrin fibers for a more accurate description of the mechanical behavior of blood clots under external forces. Maksudov et al. [11] and Filla et al. [12] have proposed different approaches to address this gap. Maksudov et al. [11]

introduced a model grounded in a superposition of probability densities, while Filla et al. [12] proposed a model based on the linear combination of a simple harmonic oscillator and a one-sided harmonic potential. Despite these advancements, the quest for hyperelastic potentials uniquely tailored for fibrin and blood clots remains challenging. There is still an imminent need to develop new models that aspire to comprehensively describe these materials, encompassing compressibility and applicability to all deformations while also demonstrating alignment with experimental data [5-8].

The present models addressing the elasticity of fibrin in blood clot formation that satisfy the specified criteria include the Purohit et al. [8] model, designed for delineating fibrin gel mechanics, and the Fereidoonnezhad-McGarry model [5-7], developed for elucidating blood clot mechanics. Notably, these models precede the Maksudov et al. [11] and Filla et al. [12] fibrin fiber models. Purohit et al. [8] characterized the fibrin fiber as a worm-like chain, deriving their fibrin gel potential from the mechanics of the Arruda-Boyce unit cell. The resultant model demonstrated remarkable accuracy in predicting tensile stress at low to moderate strains and effectively forecasted the volume change of a fibrin clot under tension. However, its accuracy diminished when predicting tensile stress at large strains. Notwithstanding this limitation, the model, grounded in first-principles derivation, establishes a meaningful connection between the predicted stress-strain behavior of fibrin clots and underlying physical and microscopic parameters.

On the other hand, the Fereidoonnezhad-McGarry model took a quite different approach [5-7]. Instead of following the work of Arruda and Boyce, the Fereidoonnezhad-McGarry model treats the blood clot as a composite material of a solid matrix embedded with fibers. To capture the well-known bilinear force-extension behavior of fibrin fibers, Fereidoonnezhad and McGarry described the fiber contribution using a piece-wise polynomial equation and separated the strain-energy density into isochoric (shape change) and volumetric (volume change) contributions. The solid matrix strain-energy (also described by a set of piece-wise equations) and the fiber strain-energy were defined to only contribute to shape changes. A separate set of piece-wise equations, describing a modified ideal gas law, were used to describe the energy changes resulting from volume changes. The Fereidoonnezhad-McGarry model outperforms the classical theories of nonlinear elasticity or the Purohit et al. model in the accuracy of describing the mechanics of blood clots. Furthermore, it can be generalized to treat isotropic and anisotropic blood clots. However, the piece-wise-based model only guarantees $C^0$ and $C^1$ continuity, meaning the strain-energy function and stress-strain functions are continuous, but the tangent-stiffness functions are not. Also, unlike the model by Purohit et al., the Fereidoonnezhad-McGarry model is purely phenomenological and based on theories that may not be physically feasible when applied to blood clots. Lastly, Fereidoonnezhad and McGarry have defined their model to be fit to experimental data as a fictitious stress-measure instead of a real stress-measure.

To overcome the limitations of the existing models, we continue the efforts of developing new constitutive models for more accurate representations of the mechanical behavior of fibrin and blood clots. First, we proposed a hyperelastic potential composed of a series of strain-energy terms to capture the various experimentally observed/deduced microscopic deformation mechanisms of blood clots, including fiber stretching [13-18], fiber buckling [19-22], fiber bending [19-22], densification of RBCs and fibrin fibers [20, 22], as well as RBC and fibrin fiber jamming [23-26]. The resulting model was validated against three independent experimental data sets: a fibrin clot in tension from Brown et al. [13], a blood clot in compression from Johnson et al. [5, 26], and a blood clot in simple shear from Sugerman et al. [27]. Finally, simulations on a microscopic blood clot model that explicitly includes fibrin fibers and RBCs were performed to investigate the relation between the phenomenological parameters of the proposed hyperelastic material model and physical quantities. This model holds potential as a valuable tool in micromechanical research, contributing to a comprehensive understanding of clot-related diseases and facilitating the development of more efficacious treatments for these conditions.

## 2. Developing Energy-Informed Phenomenological Hyperelastic Model for Blood Clot

### 2.1 Strain energy for fibrin fiber under stretch

The proposed hyperelastic potential is developed by first considering the bilinear force-extension behavior of fibrin fibers. A suitable force-extension model was proposed in our previous work by Filla et al. [12], which fits experimental force-strain curves of individual fibrin fibers with high accuracy ($R^2 > 0.99$ for seven experimental fibrin fiber force-strain curves),

$$F = k_1 \varepsilon + (k_2 - k_1) \frac{1}{2} \left[ \varepsilon + (\varepsilon - m) \operatorname{erf}\left( \frac{\sqrt{\frac{1}{2}}(\varepsilon - m)}{s} \right) + \sigma \sqrt{\frac{2}{\pi}} \exp\left( -\frac{(\varepsilon - m)^2}{2s^2} \right) - C \right] \quad (1)$$

where $F$ is the tensile force, $\varepsilon$ is the fiber strain, $k_1$ and $k_2$ are stiffness constants, $m$ and $s$ are strain constants, and $C$ is a constant of integration that enforces $F = 0$ at $\varepsilon = 0$. The curve is fit to experimental fibrin fiber force-strain data in Figure 1. The first term represents the force that results from a simple harmonic oscillator. The second term $(k_2 - k_1)\frac{1}{2}[\ldots]$ describes a one-sided harmonic potential, where the force is approximately zero when ε is approximately less than $m - 2s$ and the force is approximately linearly when ε is approximately greater than $m + 2s$. In between $m - 2s$ and $m + 2s$, the function is smooth and nonlinear.

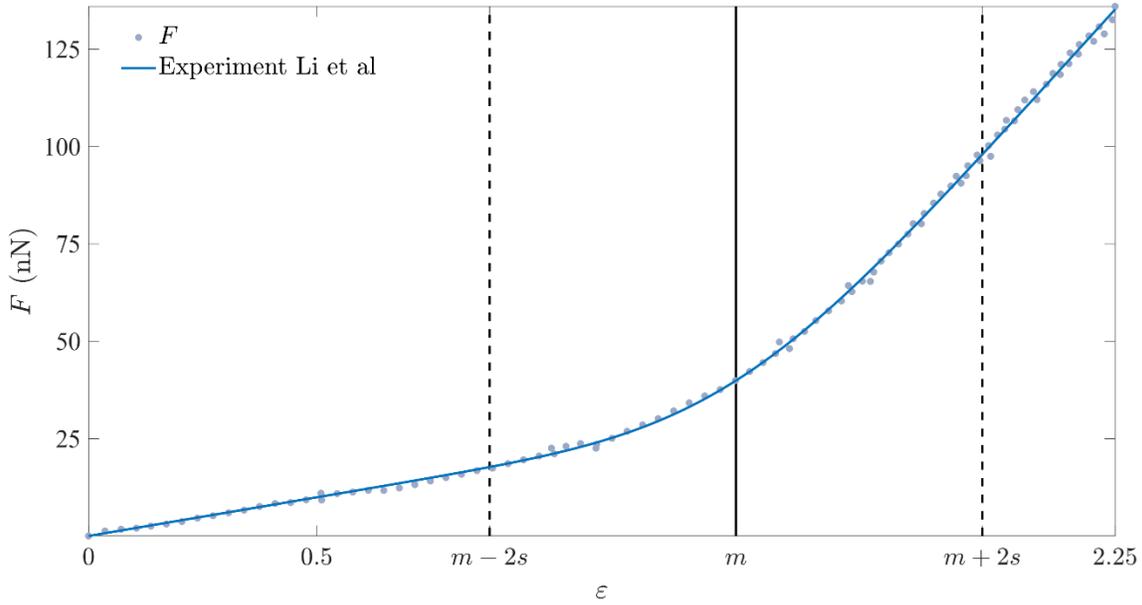

Figure 1: **Force-extension behavior of a fibrin fiber.** Measured by Li et al. [28] and regressed by the force-extension relationship given by Equation 1. Equation 1 successfully reproduces the bilinear force-strain behavior of fibrin fibers and provides a quantitative description of the transitions in the fibrin fiber force-strain curve.

The one-sided harmonic potential describes a transition from a soft linear force-extension behavior to a stiffer linear force-extension behavior. The mechanistic explanation of Equation 1 is as follow: $k_1$, is the stiffness resulting from the entropic extension of alpha C-regions in the fibrin fiber, and $k_2$ is the stiffness resulting from the enthalpic extension of unfolded protein [28-39]. The one-sided harmonic potential is defined by the integral $\int \left(1 + erf\left[\frac{\varepsilon - m}{s\sqrt{2}}\right]\right) d\varepsilon$. To incorporate this fibrin fiber behavior into the blood clot model, we start with the simplest kinematic assumptions: 1) the microscopic fiber strains are equal to the macroscopic strains, and 2) the fiber orientations are isotropically distributed. With these two assumptions the strain-energy density of the one-sided simple harmonic potential can be rewritten as a hyperelastic material model,

$$\psi_{RSH} = \frac{\mu}{2} f^*\{\lambda_i; \vartheta^*, \Delta^*\} = \frac{\mu}{2} \int_1^{\lambda_i} \int_1^y \left(1 + \text{erf}\left[\frac{x-\vartheta^*}{\Delta^*\sqrt{2}}\right]\right) dxdy, \text{ with } i = 1, 2, 3 \qquad (2)$$

where $\mu$ is the shear modulus when $\lambda_i > \vartheta^* + 3\Delta^*$, $\lambda_i$ is the $i^{th}$ principal stretch, and $\vartheta^*$ and $\Delta^*$ are material dependent parameters taking the place of $m$ and $s$. The parameter $\vartheta^*$ is equal to the isochoric stretch required to reach the halfway point of a transition from one deformation mechanism to the next one. The parameter $\Delta^*$ describes the width of the nonlinear transition region from one deformation mechanism to the next one. The explicit formula resulting from $\int_1^{\lambda_i} \int_1^y \left(1 + \text{erf}\left[\frac{x-\vartheta^*}{\Delta^*\sqrt{2}}\right]\right) dxdy$ is given in Appendix A. The subscript $RSH$ in $\psi_{RSH}$ means right-sided harmonic. Correspondingly, in the subsequent section, a left-sided harmonic potential will be introduced to elucidate the jamming behavior of clots under volumetric compression.

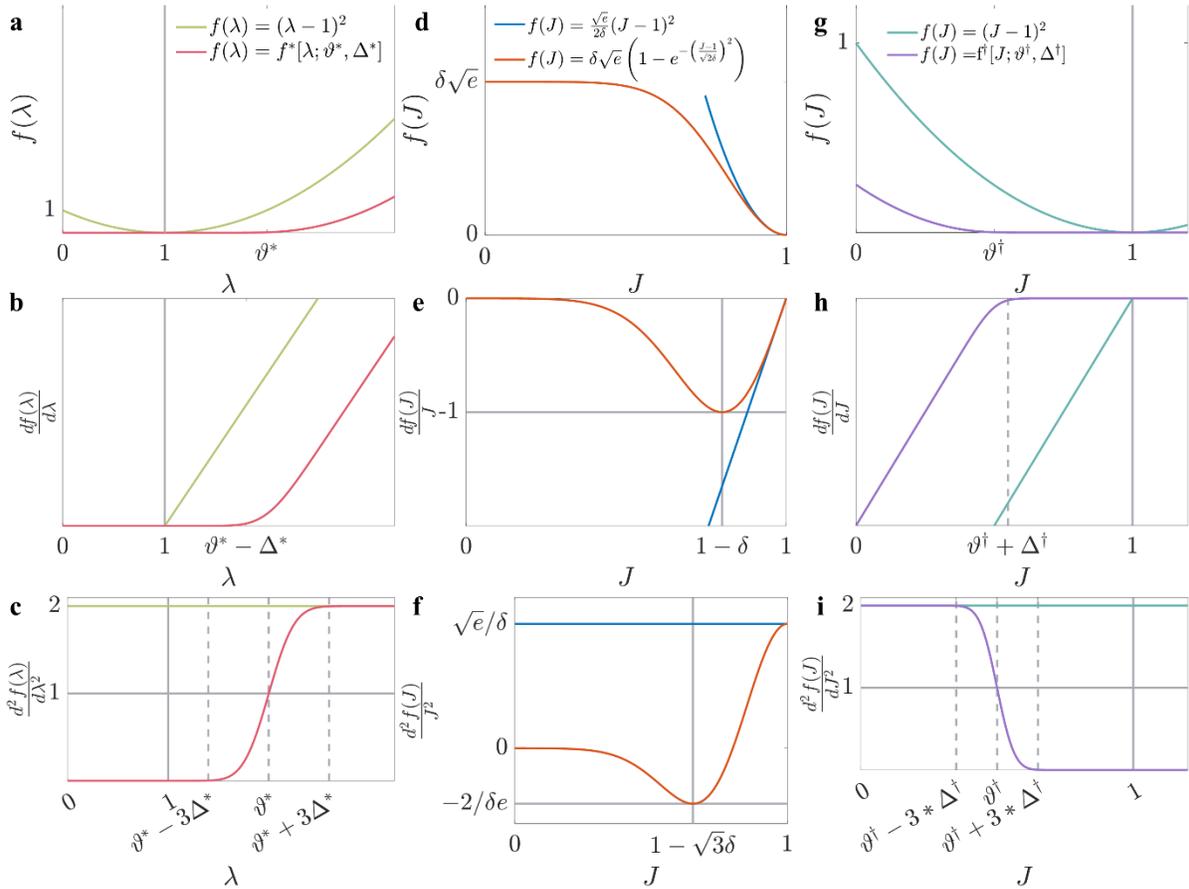

Figure 2: **Trend of key items in $\psi_{RSH}$ during fiber tension, fiber buckling, and fiber jamming**. Behavior of Equations 2, 6, and 8 describing **a-c** a right-sided harmonic potential, **d-f** a Gaussian potential, and **g-i** a left-sided harmonic potential. The right-sided harmonic potential is used to describe the various stiffness transitions of fibrin fibers in a mesh. The Gaussian potential is used to describe fiber buckling. The left-sided harmonic potential is used to describe the transition from densification to jamming in clots subjected to volumetric contraction.

The properties of $\frac{\mu}{2}f^*\{\lambda_i;\vartheta^*,\Delta^*\}$ are depicted in Figure 2a-c in comparison to a simple harmonic oscillator based on Biot strain $\frac{\mu}{2}(\lambda-1)^2$. In both cases, $\mu$ has been set equal to 2s, so the leading coefficient is equal to one for a clear comparison of the stretch-dependent terms. In Figure 2a, $f^*\{\lambda_i;\vartheta^*,\Delta^*\}$ is approximately zero until the stretch is approximately equal to $\vartheta^*$, at which point the function appears to initiate increasing. In Figure 2b, the first derivative of $f^*\{\lambda_i;\vartheta^*,\Delta^*\}$ with respect to $\lambda$ is nearly zero until $\lambda \approx \vartheta^* - \Delta^*$, at which point the function appears to increase and eventually converge to the first derivative of $(\lambda-1)^2$ (at $\lambda \approx \vartheta^* + 2\Delta^*$). Finally, Figure 2c shows that the second derivative of $f^*\{\lambda_i;\vartheta^*,\Delta^*\}$ with respect to $\lambda$ is nearly zero until $\lambda \approx \vartheta^* - 3\Delta^*$, is equal to one at $\lambda = \vartheta^*$, and is approximately equal to the second derivative of $(\lambda-1)^2$ at $\lambda \approx \vartheta^* + 3\Delta^*$. With the behavior of $f^*\{\lambda_i;\vartheta^*,\Delta^*\}$ depicted, it is clearly shown that Equation 2 describes a one-sided harmonic potential. In particular, at large enough strains (depending on $\vartheta^*$ and $\Delta^*$), the behavior of Equation 2 is nearly indistinguishable from a simple harmonic oscillator.

Before defining strain-energy terms based on Equations 1 and 2, some details about the mechanics of fibrin gels should be considered. While the fibrin fiber itself is well described as a linear combination of a simple harmonic oscillator and a one-sided harmonic potential, experimental observation suggests that the fibrin mesh would be best depicted by a linear combination of a simple harmonic oscillator and two one-sided harmonic potentials. This is intuited from low-strain tensile stress-strain data of fibrin gels, which initially exhibit a toe region at very small strains and then become linear at slightly larger strains [13-15, 17, 40]. Micrographs of fibrin gels at different strains suggest that this toe region is the result of fibers initially aligning with the applied tensile stress at very small strains [16-18], indicating that there is some small finite strain that needs to be applied to a fibrin gel before all the fibers being recruited into tension. Therefore, strain-energy terms could be proposed as,

$$\psi_{align} = \frac{\mu_a}{2}(\lambda_i - 1)^2 \tag{3}$$

$$\psi_{entropic} = \frac{1}{2}(\mu_S - \mu_a)f^*\{\lambda_i;\vartheta_S^*,\Delta_S^*\} \tag{4}$$

$$\psi_{enthalpic} = \frac{1}{2}(\mu_H - \mu_S - \mu_a)f^*\{\lambda_i;\vartheta_H^*,\Delta_H^*\} \tag{5}$$

where $i = 1, 2, 3$ for equations 3-4, $\mu_a$, $\mu_S$, and $\mu_H$ are the shear moduli due to fiber alignment, the fiber stiffness due to the entropic stretching of $\alpha C$-regions, and the fiber stiffness due to the enthalpic stretching of unfolded protein, respectively. $\vartheta_S^*$ and $\Delta_S^*$ define the transition regions from fiber aligning to fiber stretching, and $\vartheta_H^*$ and $\Delta_H^*$ define the transitions from entropic to enthalpic stretching of the nanoscopic components of individual fibrin fibers.

*2.2 Strain energy for fibrin fiber buckling*

Compression experiments have provided insight into a peculiar behavior of fibrin gels. Fibrin fibers are porous, loosely bound conglomerates of flexible fibrin protofibrils held together by long amino acid chains ($\alpha C$-regions). However, during compression, some fibrin gels exhibit a stress-strain behavior that is most intuitively described as a loading, buckling, post-buckling mechanism due to fibrin fibers [19-22], suggesting that the porous collection of flexible protein and amino acid chains can be stiff enough in proportion to their length in order to experience buckling. However, during shear experiments of fibrin gels, no such behavior is observed, suggesting that shape changes do not lead to a meaningful energy contribution from fiber buckling [15, 27, 41-46]. Therefore, a strain-energy term describing fiber buckling is also considered for this material model based on the following two assumptions: 1) energetically meaningful fiber buckling results from volume reduction, and 2) the post-buckling behavior of fibrin fibers should already be captured by $\psi_{align}$ so there is no need for a post-buckling potential. Therefore, strain-energy term for fiber buckling is proposed as,

$$\psi_{buckle} = \beta_b \delta_b \sqrt{e}\left(1 - \exp\left[-\left(\frac{J-1}{\delta_b\sqrt{2}}\right)^2\right]\right) \qquad (6)$$

where $J$ is the ratio of the deformed volume to the initial volume, $\beta_b$ is a stiffness constant, and $\delta_b$ determines at what volumetric strain the fibers have buckled. The stiffness of $\psi_{buckle}$ at small strains can be approximated by a Taylor-series expansion around $J = 1$ which results in $\frac{\beta_b\sqrt{e}}{\delta_b}(J-1)^2$, as Figure 2d shows. At $J = 1 - \delta_b$ the first derivative of $\psi_{buckle}$ reaches a minimum, the buckling point, after which the first derivative smoothly approaches zero, as shown in Figure 2e. Finally, at $J = 1$ the second derivative of $\psi_{buckle}$ is equal to $\frac{\beta_b\sqrt{e}}{\delta_b}$, then as expected, the potential predicts a negative stiffness arising from buckling that reaches a minimum at of $-\frac{2\beta_b}{\delta_b}$ at $J = 1 - \sqrt{3}\delta_b$, as shown in Figure 2f.

*2.3 Strain energy for blood clot densification and jamming*

Next, as the volume of a blood clot reduces the interstitial, RBCs are forced to change shape as the available space in the porous fiber mesh decreases [20, 22]. This energy should be proportional to volume fraction of RBCs and fibrin fibers in the clot, more specifically the total surface area of RBCs and the total volume available to the RBCs. Currently, there is not a known theoretical framework to address this phenomenon, therefore, we will consider it to follow a simple harmonic relationship as a first approximation,

$$\psi_{densification} = \beta_d (J-1)^2 \tag{7}$$

Finally, as the volume ratio $J$ of the blood clot approaches the combined volume fraction of the fibrin fibers and RBCs, $\phi$, the clot will become jammed and behave as a solid under compression [23-26]. This portion of the potential is similar to the one-sided harmonic potential, where it should be zero for $J > \phi$ and non-zero for $J \leq \phi$. The one-sided harmonic potential in Equation 2 can be characterized as right-sided, where it is zero below some particular stretch and non-zero above it. The jamming behavior is reversed and requires a left-sided harmonic potential, which is proposed as,

$$\psi_{LSH} = \frac{\beta}{2} f^{\dagger}\{J; \vartheta^{\dagger}, \Delta^{*\dagger}\} = \frac{\beta}{2} \int_1^J \int_1^y \left(1 + \text{erf}\left[\frac{x - \vartheta^{\dagger}}{\Delta^{\dagger}\sqrt{2}}\right]\right) dx dy \tag{8}$$

where $\beta$ is the bulk modulus when $J < \vartheta^{\dagger} + 3\Delta^{\dagger}$, and $\vartheta^{\dagger}$ and $\Delta^{\dagger}$ depend on material properties. The parameter $\vartheta^{\dagger}$ is equal to the volume ratio when it reaches the halfway point of a transition from one deformation mechanism to the next. The parameter $\Delta^{\dagger}$ describes the width of the nonlinear transition region from one deformation mechanism to the next. The explicit formula resulting from $\frac{\beta}{2} \int_1^J \int_1^y \left(1 + \text{erf}\left[\frac{x - \vartheta^{\dagger}}{\Delta^{\dagger}\sqrt{2}}\right]\right) dx dy$ is given in Appendix A.

The properties of $\frac{\beta}{2} f^{\dagger}\{J; \vartheta^{\dagger}, \Delta^{\dagger}\}$ are depicted in Figure 2g-i in comparison to a simple harmonic oscillator based on the volume ratio $\frac{\beta}{2}(J-1)^2$. We will not go into details about the properties of $f^{\dagger}\{J; \vartheta^{\dagger}, \Delta^{*\dagger}\}$ since they are similar to the properties of $f^*\{\lambda_i; \vartheta_S^*, \Delta_S^*\}$ described earlier. However, it should be noted that neither $f^{\dagger}\{J; \vartheta^{\dagger}, \Delta^{*\dagger}\}$ nor $f^*\{\lambda_i; \vartheta_S^*, \Delta_S^*\}$ go to infinity as $J$ or $\lambda_i$ approach to zero. Although this fact is thermodynamically incorrect, the deformations of interest for blood clots and normal materials for that matter never come close to these singularities, so this issue can practically be ignored. This issue is usually alleviated by adding additional terms, resembling the work done by expanding/contracting a one-dimensional and three-dimensional ideal gas, such as $\ln[\lambda]$ and $\ln[J]$. The proposed model has grown complicated enough without the addition of these terms, and the consequence of not including them in a material model for blood clots is arbitrary. With Equation 8, the strain-energy resulting from jamming can be written as,

$$\psi_{jammed} = \frac{\beta_j}{2} f^\dagger\{J; \vartheta_j^\dagger, \Delta_j^\dagger\} \qquad (9)$$

where $\beta_j$ is the bulk modulus of the jammed clot, $\vartheta_j^\dagger$ and $\Delta_j^\dagger$ determine the transition region from clot densification to clot jamming. With all the strain-energy terms defined the total strain-energy density can be written as,

$$\psi = \psi_{align} + \psi_{entropic} + \psi_{enthalpic} + \psi_{buckle} + \psi_{densification} + \psi_{jammed} \qquad (10)$$

The proposed model is comprised of three one-sided simple harmonic potentials, two simple harmonic oscillators, and a Gaussian potential with a total of 13 parameters. We have assumed that 1) the microscopic fiber strains are equal to the macroscopic strains, 2) the fiber orientations are isotropically distributed, 3) energetically meaningful fiber buckling results from the volume changes rather than the shape changes, 4) the post-buckling behavior of fibrin fibers should already be captured by $\psi_{align}$ so there is no need for an extra post-buckling potential, 5) the hyperelastic potential does not need to consider the event of the volume ratio $J$ or the principal stretches $\lambda$'s approaching a singularity (approaching zero). Next, the hyperelastic potential is rigorously evaluated using experimental data, specifically observing the behavior of blood and fibrin clots under tension, compression, and shear. In parallel, this potential is critically compared with the prominent hyperelastic material model for blood clots developed by Fereidoonnezhad and McGarry, providing a comprehensive assessment of its validity and effectiveness.

## 3. Evaluating and Validating Hyperelastic Blood Clot Continuum Model for Mechanical Response

To test the blood clot model, we have curated data from three experiments published in the literature: 1) tensile test of a fibrin gel including stress-strain and volume change measurements, 2) compression test of a whole blood clot including stress-strain and volume change measurements, 3) shear test of a blood clot including shear and normal stress-strain measurements. Furthermore, the predictions of the proposed model are compared to the Fereidoonnezhad-McGarry model for all tests. Fereidoonnezhad and McGarry suggested the utilization of their model involving two fiber families, which results in four piece-wise strain energy equations, each represented by three equations. This totals to twelve equations involving a combined 18 parameters. Additionally, nine more equations are required to ensure $C^0$ and $C^1$ continuity [7]. Furthermore, Fereidoonnezhad and McGarry suggested that the isochoric part of their potential should be differentiated with respect to isochoric principal stretches, resulting in a fictitious stress measure [5, 7].

*3.1 Definition of engineering stress and its relationship with deformation gradient*

For a hyperelastic potential, the tensile and compressive stress-strain can be derived directly from $P_{11} = P_1 = \frac{\partial \psi}{\partial \lambda_1}$, where $P_{11}$ is the tensile or compressive stress equal to the first principal stress $P_1$ ($P$ is the first Piola-Kirchoff stress which is equivalent to engineering stress $\frac{Force}{Initial\ Area}$). For shear, the spectral decomposition of the first Piola-Kirchoff stress, $\boldsymbol{P} = \sum_{i=1}^{3} P_i \hat{\boldsymbol{n}}_i \otimes \widehat{\boldsymbol{N}}_i$, is required to derive the shear stress. Here, $\boldsymbol{P}$ is the first Piola-Kirchoff stress tensor, $P_i = \frac{\partial \psi}{\partial \lambda_i}$ are the eigenvalues of the first Piola-Kirchoff stress tensor, $\hat{\boldsymbol{n}}_i$ are the normalized eigenvectors of the left Cauchy-Green strain tensor ($\boldsymbol{b} = \boldsymbol{FF}^T$), $\widehat{\boldsymbol{N}}_i$ are the normalized eigenvectors of the right Cauchy-Green strain tensor ($\boldsymbol{C} = \boldsymbol{F}^T\boldsymbol{F}$), $\otimes$ denotes the tensor product, and $\boldsymbol{F}$ is the deformation gradient defined as $\boldsymbol{F} = \begin{bmatrix} 1 & \gamma & 0 \\ 0 & 1 & 0 \\ 0 & 0 & 1 \end{bmatrix}$ for a simple shear deformation. Careful application of these formulas leads to $P_{12} = \frac{\lambda_1^2}{\lambda_1^2+1} \frac{\partial \psi}{\partial \lambda_1} - \frac{\lambda_2^2}{\lambda_2^2+1} \frac{\partial \psi}{\partial \lambda_2}$ and $P_{11} = \sqrt{\frac{1}{\lambda_1^2+\lambda_2^2+2}} \left(\frac{\partial \psi}{\partial \lambda_1} + \frac{\partial \psi}{\partial \lambda_2}\right)$, where $P_{12}$ and $P_{11}$ are the engineering shear stress and normal stress on the sheared face resulting from a simple shear. The derivation of tensile/compressive stress, shear stress, and normal stress during shear for this model are provided in Appendix B and C.

*3.2 Evaluation and comparison of multiple material models for hyperelastic blood clot*

The tensile stress-strain equation derived from the strain-density function of our proposed model (Equation 10) and the Fereidoonnezhad-McGarry model were regressed against the tensile stress-strain data from Brown et al. [13] (Figure 3a). During regression, the residual sum of squares (RSS) of the tensile stress $\sum(P_{11,model} - P_{11,truth})^2$ and the RSS of the normal stress on the unloaded face $\sum(P_{22,model} - 0)^2$ were minimized (Figure 3b). Values for $\lambda_2$ and $\lambda_3$ used during the regression were estimated from a phenomenological fit to the volume change data from Brown et al. [13] (Figure 3c). Both Equation 10 and the Fereidoonnezhad-McGarry model successfully describe the tensile stress $P_{11}$ measured in experiments, achieving $R^2 = 1 - \frac{RSS}{TSS}$ values of 0.9996 and 0.9980, respectively (where *TSS* is the total sum of squares). These $R^2$ values were calculated by exclusively considering the residuals of $P_{11}$. Furthermore, Akaike's Information Criterion,

($AIC = N \ln\left[\frac{RSS}{N}\right] - 2k$, where $N$ is the number of observations and $k$ is the number of model parameters), was used to compare the meaningfulness of each model (smaller is better). Equation 10 recorded an AIC value of 921.9 and the Fereidoonnezhad-McGarry model recorded an AIC value of 1031.3. The relative likelihood that the Fereidoonnezhad-McGarry model minimizes information loss is then $e^{\frac{1}{2}(921.9-1031.1)} = 1.9 \times 10^{-24}$. Likewise, while our proposed model nearly predicts zero stress on the unloaded surfaces (on the order of millipascals), the Fereidoonnezhad-McGarry model predicts large stresses on the unloaded surfaces (on the order of kilopascals), see Figure 2b. From visual inspection, our proposed model is nearly a perfect match for the measured tensile stress-strain data, while the Fereidoonnezhad-McGarry model fails to accurately predict the low to moderate stress-strain regime.

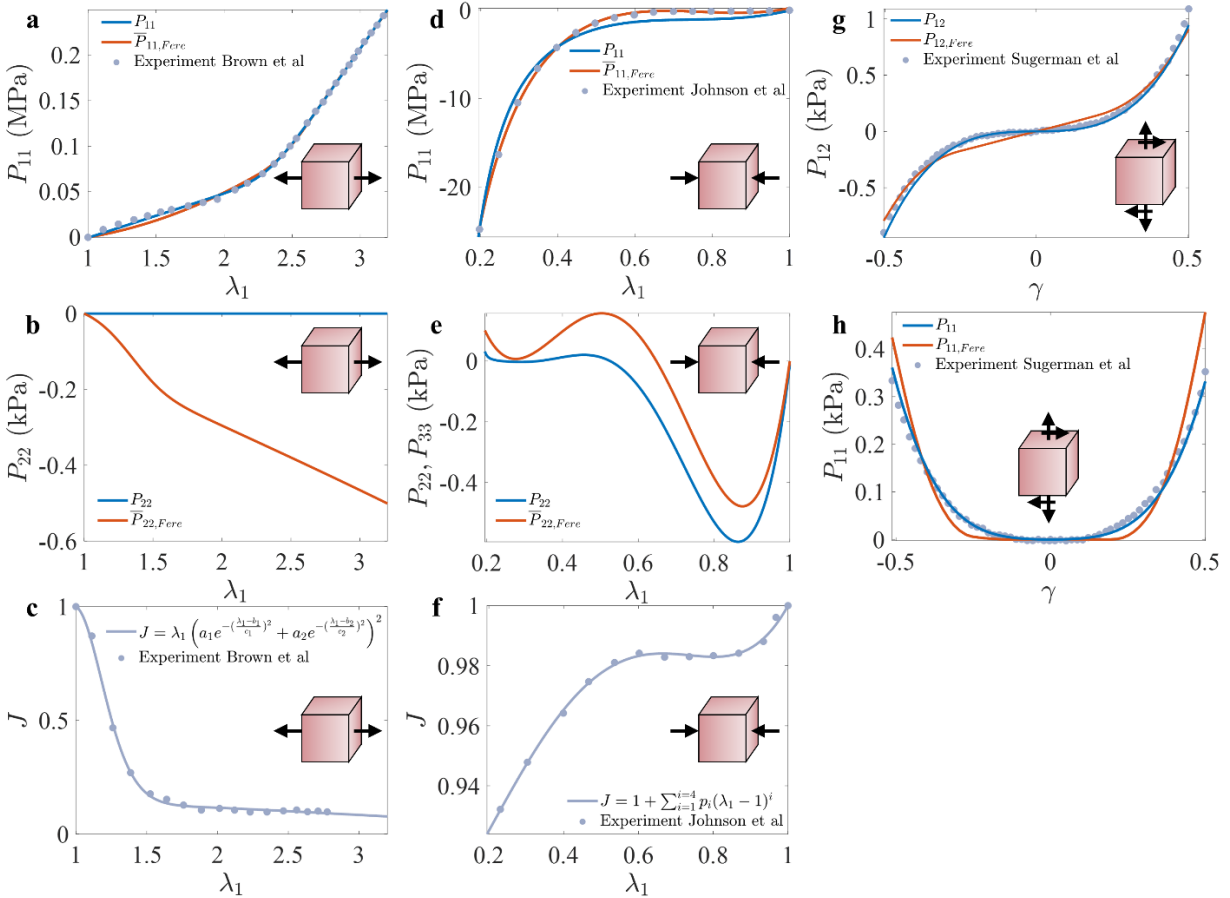

Figure 3: **Quantitative comparison of the proposed model and the Fereidoonnezhad-McGarry model**. Regression of Equation 10 in our proposed model and the Fereidoonnezhad-McGarry model against **a-c** tensile stress-strain behavior of fibrin clots [13], **d-f** compressive behavior of blood clots [26], and **g-h** shear behavior of blood clots [27].

**A-b** The tensile behavior was regressed by simultaneously minimizing the residual sum of squares of $P_{11}$ and $P_{22}$, assuming the zero-stress free-boundary condition for $P_{22}$. **c** Corresponding values for $\lambda_2$ and $\lambda_3$ were calculated from a phenomenological equation fit to $J$ vs $\lambda_1$. **d-e** The compressive behavior was regressed by simultaneously minimizing the residual sum of squares of $P_{11}$ and $P_{22}$, assuming the zero-stress free-boundary condition for $P_{22}$. **f** Corresponding values for $\lambda_2$ and $\lambda_3$ were calculated from a phenomenological equation fit to $J$ vs $\lambda_1$. **g-h** The shear behavior was regressed by simultaneously minimizing the residual sum of squares of $P_{12}$ and $P_{11}$ assuming zero volume change.

The compressive stress strain equation derived from the strain-density function of our proposed model and the Fereidoonnezhad-McGarry model were regressed against the tensile stress-strain data from Johnson et al. [26] (Figure 3d). During regression, the RSS of the tensile stress $\sum(P_{11,model} - P_{11,truth})^2$ and the RSS of the normal stress $\sum(P_{22,model} - 0)^2$ were minimized assuming zero stress along the unloaded surfaces (Figure 3e). Values for $\lambda_2$ and $\lambda_3$ used during the regression were estimated from a phenomenological fit to the volume change data from Johnson et al. [26] (Figure 3f). Both our proposed model and the Fereidoonnezhad-McGarry model successfully described the compressive stress $P_{11}$, achieving $R^2$ values of 0.9855 and 0.9995, respectively. Notably, these $R^2$ values were calculated by only considering the residuals of $P_{11}$. Furthermore, our proposed model recorded an AIC value of 254.1, while the Fereidoonnezhad-McGarry model recorded an AIC value of 207.1. These AIC scores suggest the added parameters of the Fereidoonnezhad-McGarry model are meaningful for describing the compressive behavior of blood clots. The relative likelihood that our proposed model minimizes information loss is then $e^{\frac{1}{2}(207.1-254.4)} = 5.3 \times 10^{-11}$. Both our proposed model and the Fereidoonnezhad-McGarry model predict significant normal stresses on the unloaded faces of the clot. Based on the experimental design of the blood clot compression test, the cylindrical clot was compressed between two large plates with diameters much greater than the clot. Therefore, it is reasonable that friction between the plates and clots inhibited the clot from expanding freely in the unloaded directions, which would result in the clot experiencing a significant compressive stress in the unloaded directions. As such, fitting the compressive data under the assumption of zero stress on the unloaded faces may be incorrect here. From visual inspection, the Fereidoonnezhad-McGarry model is nearly a perfect match to the measured compressive stress, while our proposed model fails to accurately predict the low to moderate stress-strain regime.

The shear stress strain equation and the normal stress during shear derived from the strain-density function of our proposed model and the Fereidoonnezhad-McGarry model were regressed against the shear stress and normal stress-strain data from Sugerman et al. [27] (Figure 3g-h).

During regression the RSS of the two models' shear stress $\sum(P_{12,model} - P_{12,truth})^2$ and the normal stress $\sum(P_{11,model} - P_{11,truth})^2$ were minimized. Both our proposed model and the Fereidoonnezhad-McGarry model reasonably describe the experimental shear stress $P_{12}$ and normal stress $P_{11}$, achieving $R^2$ values of 0.9876 and 0.9688, respectively. These $R^2$ values were determined by considering the residuals of both $P_{12}$ and $P_{11}$ simultaneously. Our proposed model recorded an AIC value of 800.4, while the Fereidoonnezhad-McGarry model recorded an AIC value of 920.5. These AIC scores suggest that the added parameters of the Fereidoonnezhad-McGarry model are not meaningful for describing the shear behavior of blood clots. The relative likelihood that the Fereidoonnezhad-McGarry model minimizes information loss is then $e^{\frac{1}{2}(800.4-920.5)} = 8.3 \times 10^{-27}$. From visual inspection, our proposed model is nearly a perfect match for the measured shear stress and normal stress, while the Fereidoonnezhad-McGarry model fails to predict the correct shape of the stress-strain curves.

Overall, our results show that the proposed model was more accurate in describing tensile and shear stress-strain data, while the Fereidoonnezhad-McGarry model was more precise in describing compression stress-strain data. Although it is beyond the scope of this study, conducting a thorough comparison of the two models by scrutinizing each energy term individually holds the potential to unveil a model that surpasses both our proposed model and the Fereidoonnezhad-McGarry model. To complete the comparison, the $R^2$ values when considering all the residuals for our proposed model and the Fereidoonnezhad-McGarry model were 0.9998 and 0.9993, respectively, while the AIC of our proposed model and the Fereidoonnezhad-McGarry model were 2296.5 and 2538.3, respectively.

## 4. Bridging the Experimental Data Gap by Computational Microscopic Model of Blood Clot

In previous sections, a hyperelastic potential was proposed for blood clots. Each strain energy term was employed to describe a unique deformation mechanism of the blood clot. However, further investigation is necessary to confirm whether these strain energy terms accurately describe the phenomena they are proclaimed to represent. Given a large enough dataset comprising stress-strain data alongside volume change data for tension and compression scenarios, as well as comprehensive microstructure details of each clot, these claims could be verified. Unfortunately, no such data set exists or may ever exist in our lifetime from experimentation.

Therefore, we have constructed a microscopic blood clot model, where the strain energy changes due to deformation can be determined through computational modeling, as the microscopic properties are entirely simulated.

*4.1 Geometric model of blood clot with microscopic components*

The model is composed of fibrin fibers treated as bead-spring fibers laid onto a perturbed cubic mesh, which is infused with RBCs treated as flexible, biconcave, surfaces that resist volume change (Figure 4). The geometry of the proposed model allows for variations in fiber mesh size $\xi$, fiber mesh tortuosity $\tau$, fiber volume fraction $\phi_f$, and RBC volume fraction $\phi_c$ (Figure 4). Fiber mesh size refers to the average length of a fiber segment between two mesh vertices (a fiber segment is colored red in Figure 4b). Fiber mesh tortuosity is calculated as the average of the shortest distances traveled from one side of the simulation box to the other following fiber segments, divided by the total length of the box (one such path is colored cyan in Figure 4b). Fiber/RBC volume fractions represent the total volumes occupied by fiber/RBCs divided by the volume of the simulation box.

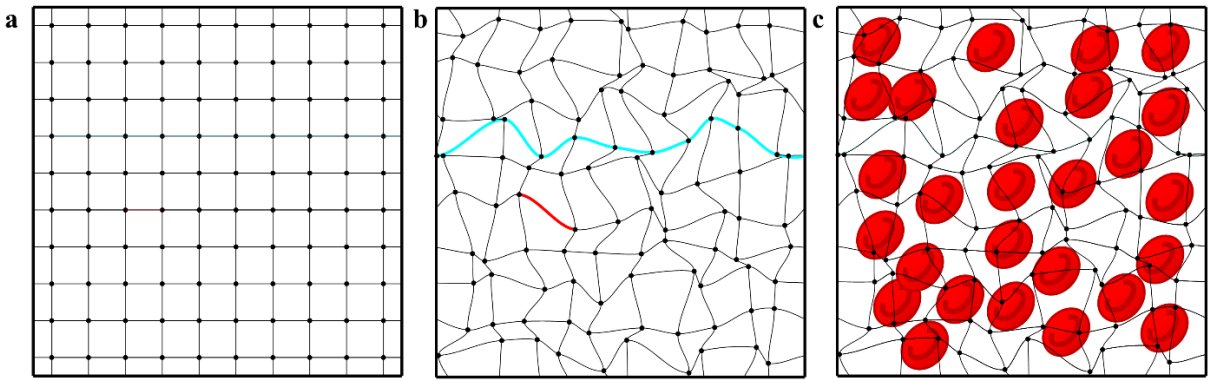

Figure 4: **A simple coarse-grained geometrical model of a fibrin mesh impregnated with RBCs**. The microscopic blood clot model geometry is modeled by the process with 2D illustrations in **a-c**. **a** A simple-cubic lattice is initialized. **b** The vertices of the simple-cubic lattice are perturbed by some small amount. After the perturbation, the mesh tortuosity is stochastic and greater than one and the mesh size is stochastic. **c** RBCs are inserted into the mesh.

*4.2 Coarse-grained potential energy of fibrin fibers and RBCs*

The energy changes for each component of the model are detailed in [47] for fibrin fibers and [48] for RBCs. The potentials developed in the aforementioned publications are described here briefly. The stretching energy of a fiber is defined as,

$$U_{T,f} = \frac{1}{2r_{ij,0}}\left(k_S(r_{ij} - r_{ij,0})^2 + r_{ij,0}^2(k_S - k_H)\frac{1}{2}\int_{r_{ij,0}}^{r_{ij}}\int_{r_{ij,0}}^{r_{ij}} 1 + \text{erf}\left(\frac{r_{ij}/r_{ij,0} - m - 1}{s\sqrt{2}}\right)dr_{ij}\right) \quad (11)$$

where $r_{ij}$ and $r_{ij,0}$ are the instantaneous and equilibrium bond length between particles, respectively, $k_S$ is the entropic stiffness of $\alpha C$-regions, $k_H$ is the enthalpic stiffness of fibrin monomers, and $\frac{1}{2}\int_{r_{ij,0}}^{r_{ij}}\int_{r_{ij,0}}^{r_{ij}} 1 + \text{erf}\left(\frac{r_{ij}/r_{ij,0} - m - 1}{s\sqrt{2}}\right)dr_{ij}$ is a function that defines smooth transitions from the initial stiffness $k_S$ to the final stiffness $k_H$ of the fibrin fiber. The smooth transition region is centered at a fiber strain of $m$ and has a width $\sim 3s$. $U_{T,f}$ is summed over all bonded pairs of fiber particles. Correspondingly, the bending energy of a fiber is defined as,

$$U_{b,f} = k_{b,f}(1 + \cos\theta_{ijk}) \quad (12)$$

where $\theta_{ijk}$ is the instantaneous angle created by three consecutively bonded particles and $k_{b,f}$ is the bending stiffness of a small segment ($2r_{ij,0}$) of the fiber. $U_{b,f}$ is summed over all fiber particles, forming a consecutively bonded chain of three particles.

The RBC model has four energy contributions to describe in-plane stretching/compressing, membrane bending, area expansion/contraction, and volume expansion/contraction. The in-plane energy is given as,

$$U_{T,c} = \frac{k_B T l_m(3x_{ij}^3 - 2x_{ij}^3)}{4p(1-x_{ij})} + \frac{k_p}{(n-1)r_{ij}^{n-1}} \quad (13)$$

where $r_{ij}$ is the instantaneous distance between two bonded particles, $l_m$ is the maximum extension of a bond, $x_{ij}$ is $\frac{r_{ij}}{l_{ij}}$, $p$ is the persistence length of the bond, $k_B T$ is Boltzmann's constant times temperature (thermal energy), $k_p$ is a spring constant determining the repulsive strength between two bonded particles, and $n$ is a power that also determines the repulsive strength between two bonded particles. $U_{T,c}$ is summed over all bonded pairs of RBC particles. The membrane bending energy is given by,

$$U_{b,c} = k_{b,c}(1 - \cos[\theta_{ijk} - \theta_0]) \quad (14)$$

where $\theta_{ijk}$ is the instantaneous angle between three consecutively bonded particles, $\theta_0$ is the equilibrium angle between three consecutively bonded particles, and $k_{b,c}$ is a bending stiffness constant. $U_{b,c}$ is summed over all RBC particles, forming a consecutively bonded chain of three particles. The energy due to area changes is separated into a global and local contribution,

$$U_A = k_A \frac{(A-A_0)^2}{2A_0} \tag{15}$$

$$U_a = k_a \frac{(a_{ijk}-a_{ijk,0})^2}{2a_{ijk,0}} \tag{16}$$

where $U_A$ is the energy change due to global area change of the entire membrane, $k_A$ is a stiffness constant, $A$ is the instantaneous area of the entire membrane, $A_0$ is the equilibrium area of the entire membrane, $U_a$ is the energy change due to local area changes, $k_a$ is a stiffness constant, $a_{ijk}$ is the instantaneous area formed by three bonded particles forming one triangle of the membrane, and $a_{ijk,0}$ is the equilibrium area formed by three bonded particles in one triangle mesh of the membrane. $U_a$ is summed over all triangles formed by bonded RBC particles. The energy due to volume changes is given by,

$$U_v = k_v \frac{(V-V_0)^2}{2V_0} \tag{17}$$

where $V$ is the instantaneous volume of the RBC, $V_0$ is the equilibrium volume of the RBC, and $k_v$ is a stiffness constant.

The potentials presented from Equations 11 to Equation 17 have been validated against experimental data for the deformation of fibrin gels and RBCs. The following potentials to describe contact between fibers and RBCs are included to enforce volume exclusion, but have not been validated against any experimental data. To date, there appears no experimental data that could be used to define contact potentials for this model. Therefore, the fiber-fiber, fiber-RBC, and RBC-RBC contact energies are determined using Hertz contact theory for spheres,

$$U_h = \frac{8}{15} E_{eff} \sqrt{R_{eff}} (R_i + R_j - R_{ij})^{5/2} \tag{18}$$

where the effective elastic modulus, $E_{eff}$, characterizing the interaction between two particles $i$ and $j$, is computed as $\left(\frac{1-v_i}{E_i} + \frac{1-v_j}{E_j}\right)^{-1}$, the effective radius, $R_{eff}$, of particles $i$ and $j$, is determined by $\frac{R_i R_j}{R_i+R_j}$, where $R_i$ and $R_j$ denote the radii of particles $i$ and $j$; $R_{ij}$ represents the distance between the two non-bonded particles.

The RBC potential is employed to describe healthy RBCs, thus, the stiffness constants are consistent to their values given in [48] during this study. On the other hand, even in healthy individuals, the microstructure of fibrin fibers can vary significantly. Therefore, additional five parameters are added into the fibrin fiber potential to refine the microscopic blood clot model. We considered the entropic fiber stiffness, $k_S$, in the range of 10 to 50 nN, the entropic fiber stiffness, $k_H$, in the range of 60 to 140 nN, the fiber strain parameter $m$ in the range of 0.8 to 1.2, the fiber strain parameter $s$ in the range of 0.1 to 0.3, and the fiber bending stiffness, $k_{b,f} = \frac{2B_s}{r_{ij,0}}$, where $B_s$ is the fiber bending stiffness, in the range of $10^{-27}$ to $10^{-23}$ Nm². The fiber bending stiffnesses studied correspond to a fiber persistence length of 1 $\mu$m to 10 mm at a body temperature of 310 K. The elastic moduli and Poisson's ratio used to calculate the Hertz potential remain constant at 1 MPa and 0.3, respectively.

*4.3 Simulation setup and preliminary analysis for tension and compression of blood clot*

The deformation of the microscopic blood clot model is conducted by LAMMPS (Large-scale Atomic/Molecular Massively Parallel Simulator) [49] and images of the clot are prepared in OVITO (Open Visualization Tool) [50]. Periodic boundary conditions were used for all simulations, and particle positions were unwrapped in OVITO to produce more intuitive images of the deformation processes. Lennard-Jones units were used to avoid division and multiplication of extremely small numbers. In the simulations, the length unit is 1 $\mu$m, the time unit is 0.0027 s, and the mass unit is 0.342 pico-kg. A timestep of $1 \times 10^{-4}$ was used during the simulations, and the mass of the fibrin and RBC particles were set to 1. Two deformation modes were used to study the blood clot model mechanics: 1) isochoric tension/compression and 2) volumetric contraction. These two deformation modes were chosen to isolate the shape change and volumetric change energy profiles, whiling avoiding the complexity introduced by spontaneous volume changes in the model. Isochoric compression can also be viewed as isochoric biaxial tension, the distinction between them is arbitrary.

Figure 5 shows images of a coarse-grained blood clot model at equilibrium, in isochoric tension, isochoric compression, and volumetric contraction. The transition of RBCs into polyhedrocytes can be observed in the volumetric compression case. However, the polyhedral shapes were less regular than those shown in experiments reported in [24-26]. During simulation, the total energy and length of the box in $x$, $y$, and $z$ directions were recorded, and the stress and tangent moduli were calculated by numerical differentiation of the strain-energy density data. Each strain-energy density curve was regressed by Equation 10 in our proposed model, allowing us to identify correlations between the microscopic properties of the blood clot model and the stiffness/strain constants of the proposed hyperelastic potential.

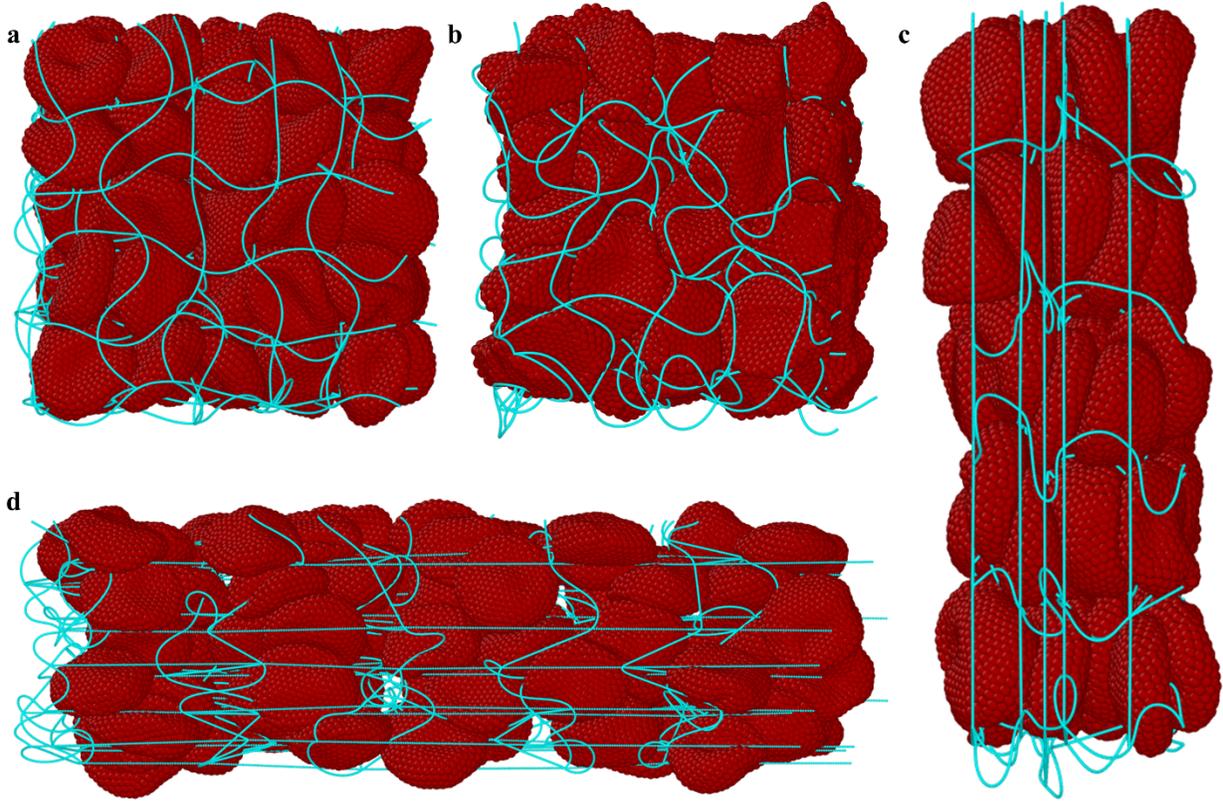

Figure 5: **Coarse-grained blood clot models under tension and compression**. **a-d** Images of the microscopic blood clot model. A microscopic blood clot model with fibers and RBCs **a** after energy minimization, **b** after being subjected to volumetric contraction, **c** after being subjected to isochoric compression, and **d** after being subjected to isochoric tension.

## 5. Building a Connection between the Continuum Description and Microscopic Mechanism of Blood Clot

In this section, a single variate analysis is conducted for the following microscopic properties of the simulated blood clots, namely the RBC volume fraction $\phi_c$, the fibrin network mesh size $\xi$, the fibrin network tortuosity $\tau$, the entropic stiffness of fibrin fibers $k_S$, the enthalpic stiffness of fibrin fibers $k_H$, and the fiber bending stiffness $B_s$. The fibrin fiber strain parameters $m$ and $s$ were varied between single variate analyses, and the fibrin fiber volume fraction changes with $\xi$, so it does not need to be explicitly studied. Among single variate analyses, randomization was applied to the variables held constant, enabling amalgamation of datasets for a robust multivariate analysis in the study's culmination. The single variate analysis helped identify direct correlations, while the multivariate analysis helped to find combined correlations between the microscopic properties of the blood clot models and the macroscopic stiffness/strain parameters of our proposed model. Some variables such as fiber network mesh size, fiber network tortuosity, and RBC volume fraction are stochastic in this model and therefore could not always be held constant, posing challenge in identifying correlations from single variate analysis. However, the multivariate analysis introduced in this section was improved by these variations.

*5.1 Effect of RBC volume fraction $\phi_c$*

The RBC volume fraction was varied from ~15% to ~70% of the total simulation box volume. The remaining adjustable parameters were held constant at $\xi \approx 6.67$ μm, $\tau \approx 1.17$, $k_S = 23$ nN, $k_H = 130$ nN, $m = 1.2$, and $s = 0.18$. The strain-energy density $\psi_{iso}$, tensile/compressive stress $P_{11}$, and tangent elastic modulus $E_t$ of the blood clot models subjected to isochoric tension and compression are shown in Figure 6a-c. The hyperelastic potential provided by our model successfully fit these strain-energy density curves, yielding $R^2$ values ranging from 0.9961 to 0.9999. Additionally, no discernible trend was observed between the stiffness/strain constants and changes in $\phi_c$ with respect to these shape-changing deformations, suggesting RBCs do not contribute to the isochoric response of blood clots during deformation.

The strain-energy density $\psi_{vol}$, hydrostatic stress $P_h$, and tangent volume modulus $K_t$ of the blood clot models subjected to volumetric contraction are shown in Figure 6d-f. The hyperelastic potential given by proposed model fit these strain-energy density curves with $R^2$ values ranging from 0.9930 to 0.9998. The effect of RBC volume fraction is clearly demonstrated in the volumetric strain-energy density curves. As expected, as RBC volume fraction increases, the initial soft response to volumetric contraction intensifies. This is because more RBCs undergo shape changes due to the increased confinement of the contracting clot. Additionally, strain stiffening occurs at higher volume ratios $J$ because less volumetric strain is required to transition the clots into the jammed phase.

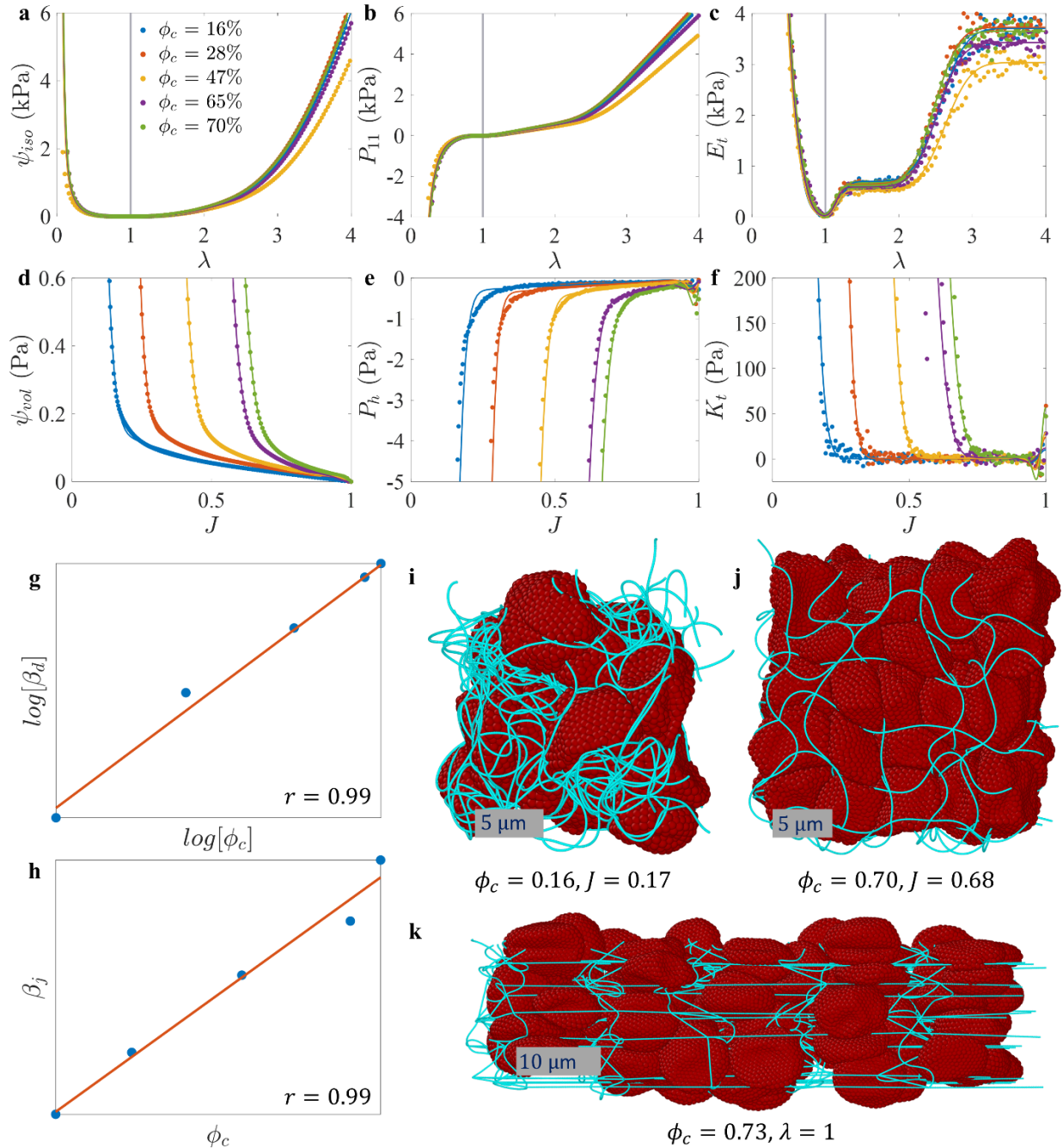

Figure 6: **Effect of RBC volume fraction**. **a-c** Strain-energy density, stress, and tangent modulus data for microscopic blood clot models having different volume fractions of RBCs in isochoric tension/compression. The strain-energy data were regressed by Equation 10 in our proposed model and no notable trends were observed. **d-f** Strain-energy density, stress, and tangent modulus data from microscopic blood clot models having different volume fractions of RBCs in volumetric contraction. The strain-energy data were regressed by Equation 10 in our proposed model and significant trends were observed. **g-h** The trends found in d-f are plotted showing a correlation between RBC volume fraction and blood clot stiffness during densification and jamming. **i-j** Microscopic blood clot models with different RBC volume fractions in the jamming phase. **k** Microscopic blood clot model with a high RBC volume fraction in isochoric tension. The RBCs make no significant contribution to energy changes in this deformation mode.

Figure 6g-h show the correlations between the RBC volume fraction and the stiffness constants of Equation 10 in our proposed model. The stiffness constant, $\beta_d$, defined in Equation 7 to evaluate the strain-energy contribution due to blood clot densification, is log-log proportional to $\phi_c$ with a Pearson correlation coefficient of 0.99. This proportionality suggests $\beta_d$ is a power-law of $\phi_c$. The stiffness constant, $\beta_j$, describing the jammed stiffness of the blood clot in Equation 9, is linearly proportional to $\phi_c$ with a Pearson correlation coefficient of 0.99. Figure 6i-k present snapshots of the simulations used for this analysis. The highlighted moments show the contrast between a blood clot with a low- and high-volume fraction of RBCs in the jammed phase, illustrating how RBCs easily traverse through the clot when the clot's shape changes while maintaining a constant volume.

*5.2 Effect of fibrin network mesh size $\xi$*

The fibrin network mesh size ranges from 6.5 μm to 8.5 μm. Notably, these mesh sizes are relatively large for blood clots. However, attempting to model smaller mesh sizes posed a significant geometrical challenge in terms of inserting RBCs without encountering issues such as fibers penetrating the RBC volumes. The remaining adjustable parameters were held constant at $\phi_c \approx 0.72$, $\tau \approx 1.10$, $k_S = 33$ nN, $k_H = 123$ nN, $m = 0.96$, and $s = 0.26$. The strain-energy density $\psi_{iso}$, tensile/compressive stress $P_{11}$, and tangent elastic modulus $E_t$ of the blood clot models subjected to isochoric tension and compression are shown in Figure 7a-c. The hyperelastic potential given by in our proposed model fit these strain-energy density curves with $R^2$ values ranging from 0.9992 to 0.9997. The effect of mesh size, $\xi$, on the isochoric strain energy density, $\psi_{iso}$, is readily discernible, particularly in the tangent modulus plot. Figure 7c shows that, as mesh size decreased, the stiffness of the blood clots in isochoric tension is enhanced for all stretches. While this trend also holds for isochoric compression, it may not be as readily apparent in these plots.

The strain-energy density $\psi_{vol}$, hydrostatic stress $P_h$, and tangent volume modulus $K_t$ of the blood clot models subjected to volumetric contraction are shown in Figure 7d-f. The hyperelastic potential given by our proposed model fit these strain-energy density curves with $R^2$ values between 0.9980 and 0.9988. The fibrin network mesh size, $\xi$, appears to increase the energy required to deform the clots in volumetric contraction during the densification phase.

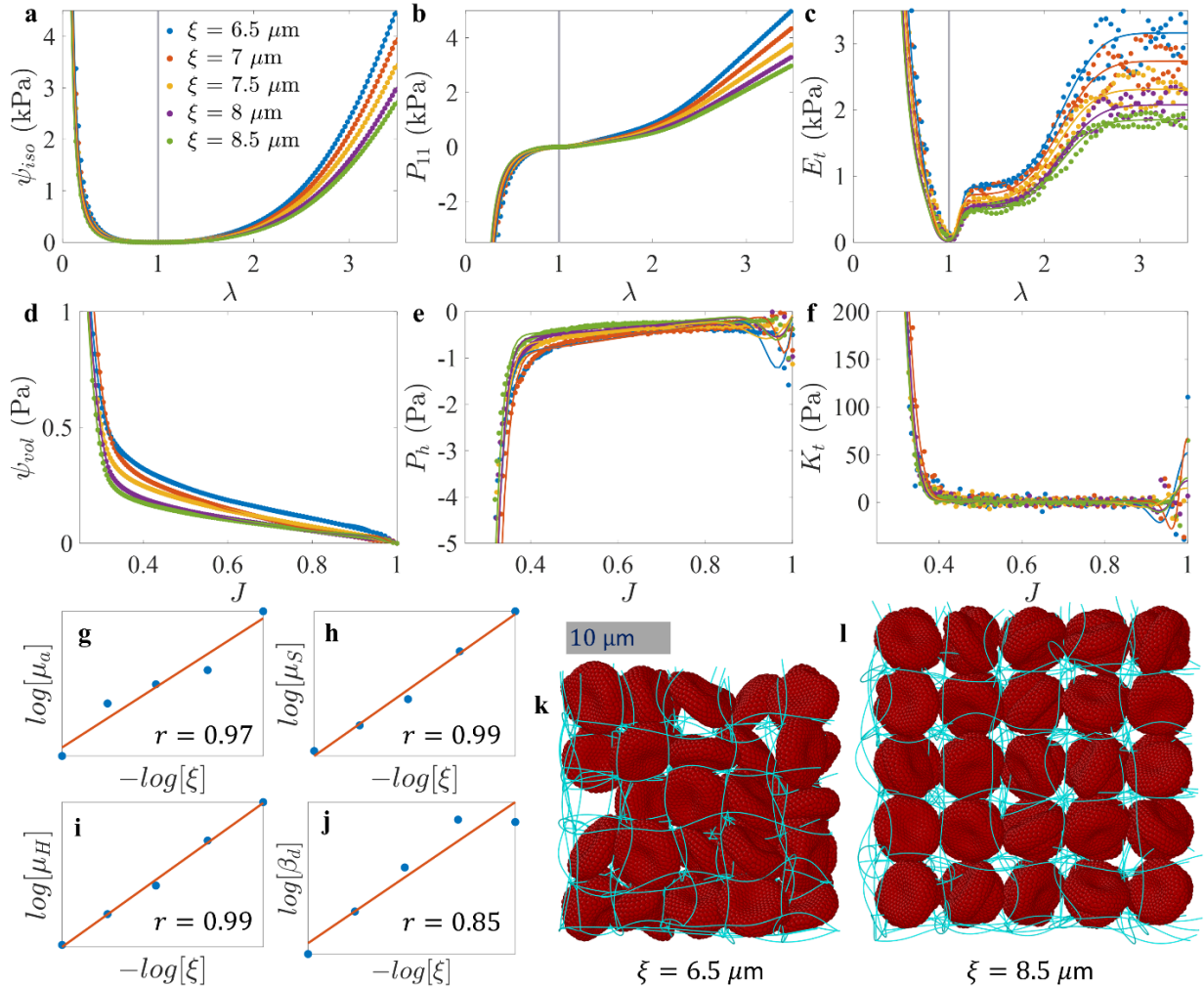

Figure 7: **Effect of fiber mesh size**. **a-c** Strain-energy density, stress, and tangent modulus data for microscopic blood clot models having different mesh sizes in isochoric tension/compression. The strain-energy data were regressed by Equation 10 in our proposed model and notable trends were identified. **d-f** Strain-energy density, stress, and tangent modulus data from microscopic blood clot models having different mesh sizes in volumetric contraction. The strain-energy data were regressed by Equation 10 our proposed model and notable trends were identified. **g-j** The trends found in a-f are plotted showing correlations between mesh size and blood clot stiffness due to fiber alignment, fiber entropic stiffness, fiber enthalpic stiffness, and fiber buckling stiffness. **k-l** Microscopic blood clot models with different mesh sizes at equilibrium showing the key distinction leading to different strain-energy curves; smaller mesh size leads to more fibers per unit volume.

Figure 7g-j show the correlations that were found between the fiber network mesh size, $\xi$, and the stiffness constants of Equation 10. The stiffness constants, $\mu_a$, $\mu_S$, $\mu_H$, and $\beta_b$ from Equations 3-6 describing the strain-energy contribution due fiber alignment (bending), entropic stiffness, enthalpic stiffness, and buckling were found to be log-log proportional to, $\xi$, with a set of Pearson correlation coefficients of 0.97, 0.99, 0.99, and 0.85 respectively. This proportionality suggests all the stiffness constants related to the fibrin fiber mesh are inverse power-laws of $\xi$.

Although the volumetric strain-energy density was affected by $\xi$, these changes appear to be captured by $\mu_a$. This suggests that fiber bending is a significant energy contribution during blood clot densification. Figure 7k-l depict snapshots of the initial configurations of blood clot models with mesh sizes of 6.5 μm and 8.5 μm. The two initial configurations show the simple explanation for the identified trends, smaller mesh sizes result in more fibers per unit volume.

*5.3 Effect of fiber mesh tortuosity $\tau$*

The fiber mesh tortuosity varies from 1.02 to 1.15. The remaining adjustable parameters were held constant at $\xi \approx 6.3$, $k_S = 13$ nN, $k_H = 64$ nN, $m = 1.06$, and $s = 0.09$. Due to the stochastic nature of the blood clot model, the RBC volume fraction was not able to be held constant during this study. This resulted from the volume change of the blood clot models during energy minimization. Changing fiber mesh tortuosity led to large changes in equilibrium volume. As a result, the RBC volume fraction ranges from 51% to 73% here and the single variate analysis is limited by this issue. The strain-energy density $\psi_{iso}$, tensile/compressive stress $P_{11}$, and tangent elastic modulus $E_t$ of the blood clot models subjected to isochoric tension and compression are shown in Figure 8a-c. The hyperelastic potential given by our proposed model fits these strain-energy density curves with $R^2$ values between 0.9755 and 0.9999. The effect of mesh tortuosity, $\tau$, on the isochoric strain energy density, $\psi_{iso}$, is not distinguishable to the naked eye. From visual inspection there appears to be no trend, however single variate analysis will show there is a measurable effect of $\tau$ on strain-energy terms that contribute to stress during shape changes.

The strain-energy density $\psi_{vol}$, hydrostatic stress $P_h$, and tangent volume modulus $K_t$ of the blood clot models subjected to volumetric contraction are shown in Figure 8d-f. The hyperelastic potential given by our proposed model fit these strain-energy density curves with $R^2$ values between 0.9992 and 0.9997. The mesh tortuosity, $\tau$, appears to have a significant effect on the volumetric strain-energy density. However, we believe this was caused by the stochastic nature of the blood clot model. Increasing mesh tortuosity resulted in a reduction in the initial volume of the clot following energy minimization. As a result, the analysis indicates that RBC volume fraction increases with mesh tortuosity, leading to similar observations as those observed in the section covering the effect of $\phi_c$. Later, the multivariate analysis will show there is no effect of fiber tortuosity on the volumetric strain-energy terms of Equation 10.

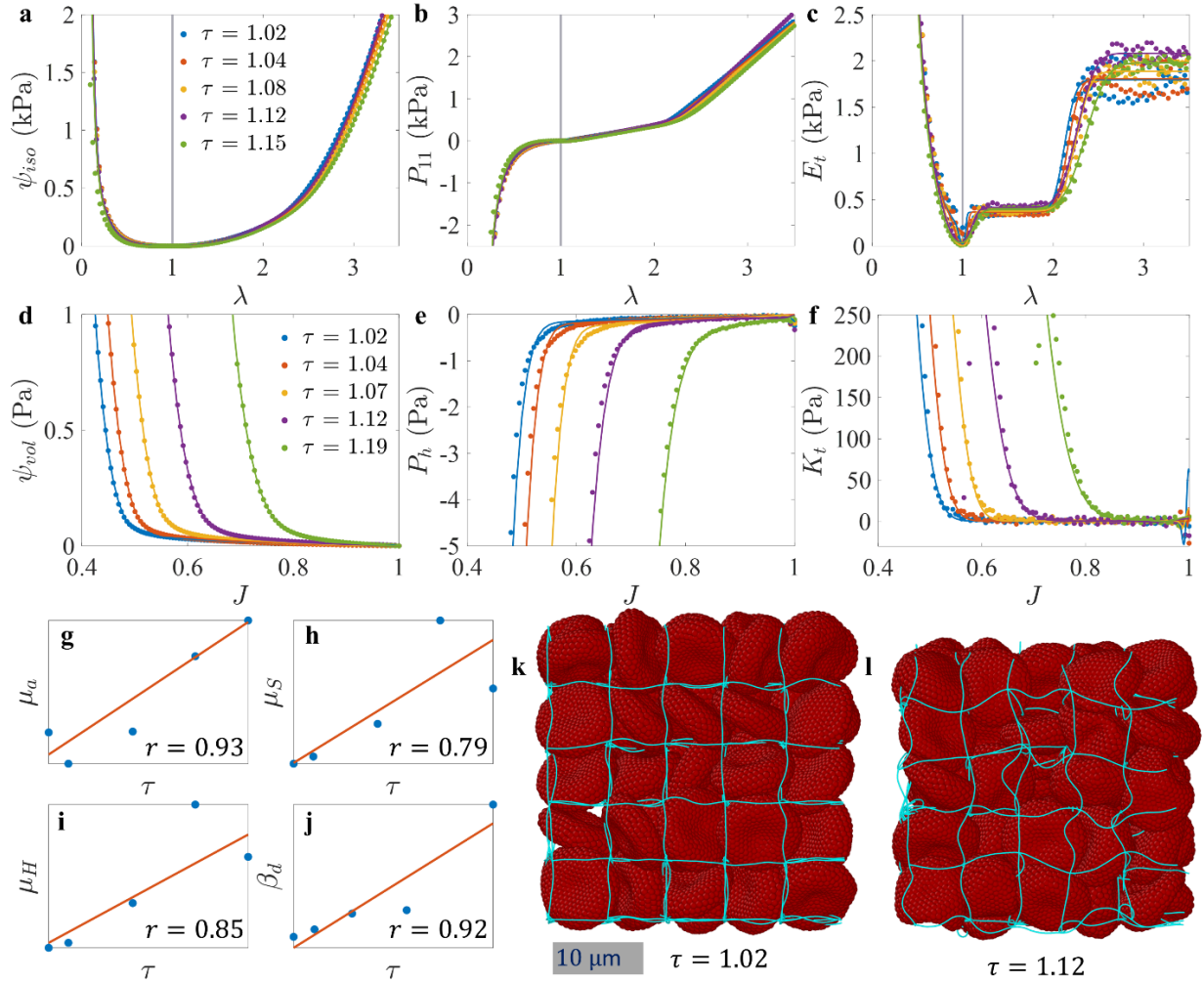

Figure 8: **Effect of fiber mesh tortuosity**. **a-c** Strain-energy density, stress, and tangent modulus data for microscopic blood clot models having different mesh tortuosity in isochoric tension/compression. The strain-energy data were regressed by Equation 10 in our proposed model and notable trends were identified. **d-f** Strain-energy density, stress, and tangent modulus data from microscopic blood clot models having different mesh tortuosities in volumetric contraction. The strain-energy data were regressed by Equation 10 in our proposed model and notable trends were identified. **g-j** The trends found in **a-f** are plotted showing a correlation between mesh tortuosity and blood clot stiffness due to fiber alignment, fiber entropic stiffness, fiber enthalpic stiffness, and fiber buckling. **k-l** Pictures of microscopic blood clot models with different mesh tortuosities at equilibrium showing the key distinction leading to different strain-energy curves; higher tortuosity leads to more fibers per unit volume.

Figure 8g-i show the correlations between the fiber mesh tortuosity, $\tau$, and the stiffness constants of Equation 10 (ignoring any correlations that would have been found from the volumetric strain energy terms). The stiffness constants, $\mu_a$, $\mu_S$, and $\mu_H$ from Equations 3-5 describing the strain-energy contribution due to fiber alignment (bending), entropic stiffness, and enthalpic stiffness were found to be directly proportional to $\tau$, with Pearson correlation coefficients of 0.93, 0.79, and 0.85, respectively. Figure 8k-l depict snapshots of the initial configurations of blood clot models with mesh tortuosity of 1.02 and 1.12. The two initial configurations show the simple explanation for the identified trends, larger mesh tortuosity results in more fibers per unit volume at equilibrium.

*5.4 Effect of fibrin fiber entropic stiffness $\mu_S$*

The fibrin fiber entropic stiffness, $k_S$, was varied from 10 nN to 50 nN. The remaining adjustable parameters were held constant at $\phi_c \approx 0.65$, $\xi \approx 6.58$, $\tau = 1.08$ nN, $k_H = 145$ nN, $m = 0.99$, and $s = 0.29$. The strain-energy density $\psi_{iso}$, tensile/compressive stress $P_{11}$, and tangent elastic modulus $E_t$ of the blood clot models subjected to isochoric tension and compression are shown in Figure 9a-c. The hyperelastic potential given by our proposed model fit these strain-energy density curves with $R^2$ values varying between 0.9998 and 0.9999. The effect of fibrin fiber entropic stiffness, $k_S$, on the isochoric strain energy density, $\psi_{iso}$, is clearly distinguishable. The effect is most easily seen in the tangent modulus plot. In Figure 9c, the stiffness at low to moderate stretches of the blood clots increases with increasing $k_S$, but the large stretch stiffness is unaffected.

The strain-energy density $\psi_{vol}$, hydrostatic stress $P_h$, and tangent volume modulus $K_t$ of the blood clot models subjected to volumetric contraction are shown in Figure 9d-f. The hyperelastic potential given by our proposed model fit these strain-energy density curves with $R^2$ values between 0.9996 and 0.9999. The fibrin fiber entropic stiffness, $k_S$, has no discernable effect on the volumetric strain-energy density, similar situation also happened to hydrostatic stress and tangent volume modulus, except for the beginning of compression. At this point, $k_S$ has little influence on $P_h$ and $K_t$. As the compression continues, the clots expand along the normal plane of the compression direction and cause stretch on fibers in normal direction. Since the entropy stretch is only a "transition" state, the fibers will finally transit from entropic stretch into enthalpic stretch. As the stretch continues, the count of fiber under entropic stretch has a trend to decrease, however, this transition is not simultaneously in all fibers, therefore, the effect from entropic stiffness is reduced and difficult to discern.

Figure 9g shows the correlation that was found between the fibrin fiber entropic stiffness $k_S$ and the stiffness constants $\mu_S$. The stiffness constant $\mu_S$ from Equation 4 describing the strain-energy contribution due to fibrin fiber entropic stiffness was found to be directly proportional to $k_S$, with a Pearson correlation coefficient of 0.99. Figure 9h depicts a blood clot at a stretch where the entropic stiffness is relevant.

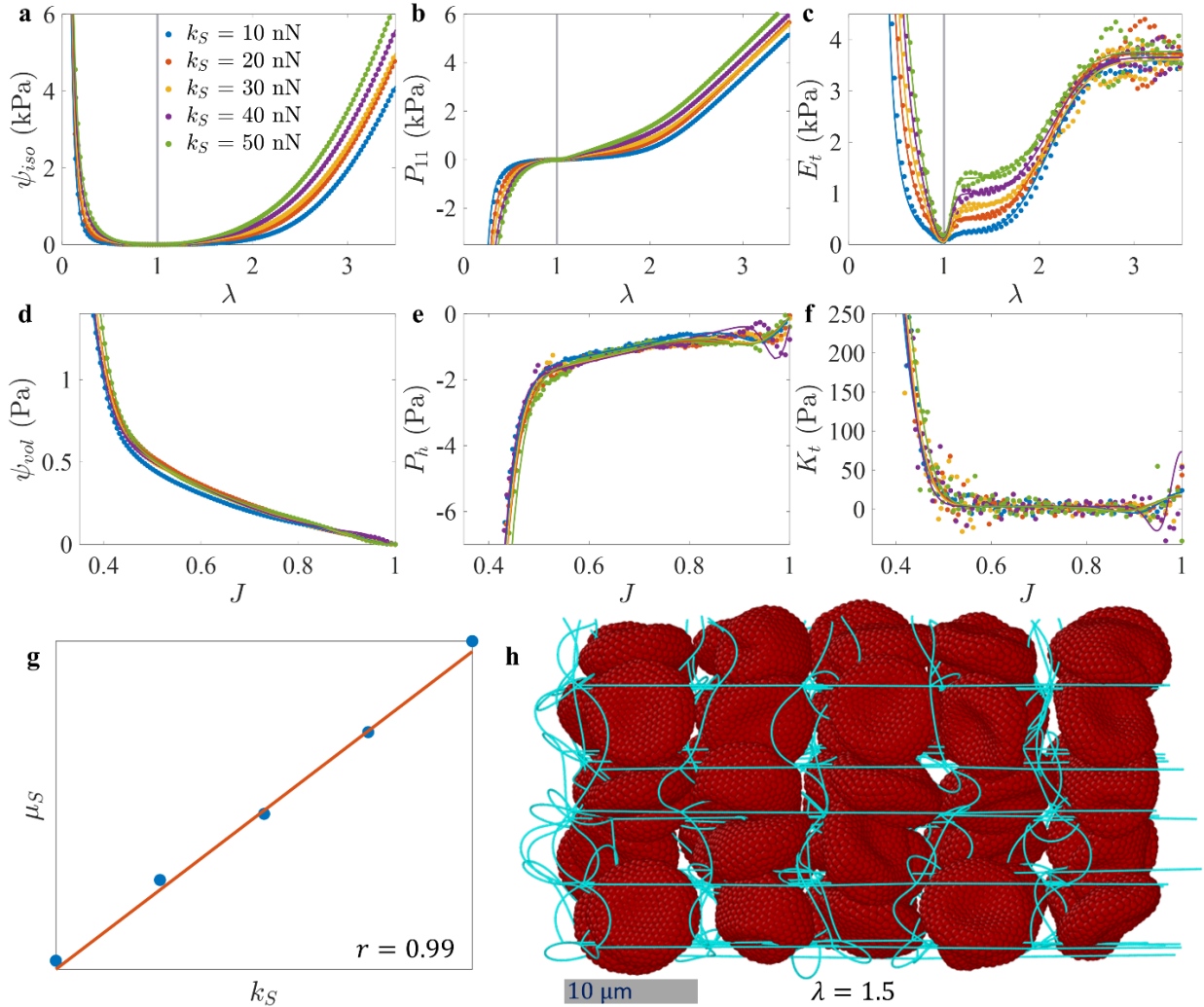

Figure 9: **Effect of fibrin fiber entropic stiffness**. **a-c** Strain-energy density, stress, and tangent modulus data for microscopic blood clot models having different fibrin fiber entropic stiffnesses in isochoric tension/compression. The strain-energy data were regressed by Equation 10 in our proposed model and one notable trend was identified. **d-f** Strain-energy density, stress, and tangent modulus data from microscopic blood clot models having different fibrin fiber entropic stiffnesses in volumetric contraction. The strain-energy data were regressed by Equation 10 in our proposed model and no notable trends were observed. **g** The trend found in a-c is plotted showing a correlation between fibrin fiber entropic stiffness and the blood clot stiffness term introduced to capture fibrin fiber entropic stiffness. **h** A pictures of a microscopic blood clot model is pictured at $\lambda = 1.5$ where the fibrin fiber entropic stiffness determines the clot stiffness.

## 5.5 Effect of fibrin fiber enthalpic stiffness $\mu_H$

The fibrin fiber enthalpic stiffness, $k_H$, was varied from 60 nN to 140 nN. The remaining adjustable parameters were held constant at $\phi_c \approx 0.51$, $\xi \approx 7.29$, $\tau = 1.15$ nN, $k_S = 48$ nN, $m = 1.23$, and $s = 0.23$. The strain-energy density $\psi_{iso}$, tensile/compressive stress $P_{11}$, and tangent elastic modulus $E_t$ of the blood clot models subjected to isochoric tension and compression are shown in Figure 10a-c. The hyperelastic potential given by our proposed model fit these strain-energy density curves with $R^2$ values ranging from 0.9090 to 0.9999. The effect of fibrin fiber enthalpic stiffness, $k_H$, on the isochoric strain energy density, $\psi_{iso}$, is clearly distinguishable. The effect is most easily seen in the tangent modulus plot. In Figure 10c, the stiffness at large stretches of the blood clots increases with augmented $k_h$, but the low to moderate stretch stiffness is unaffected.

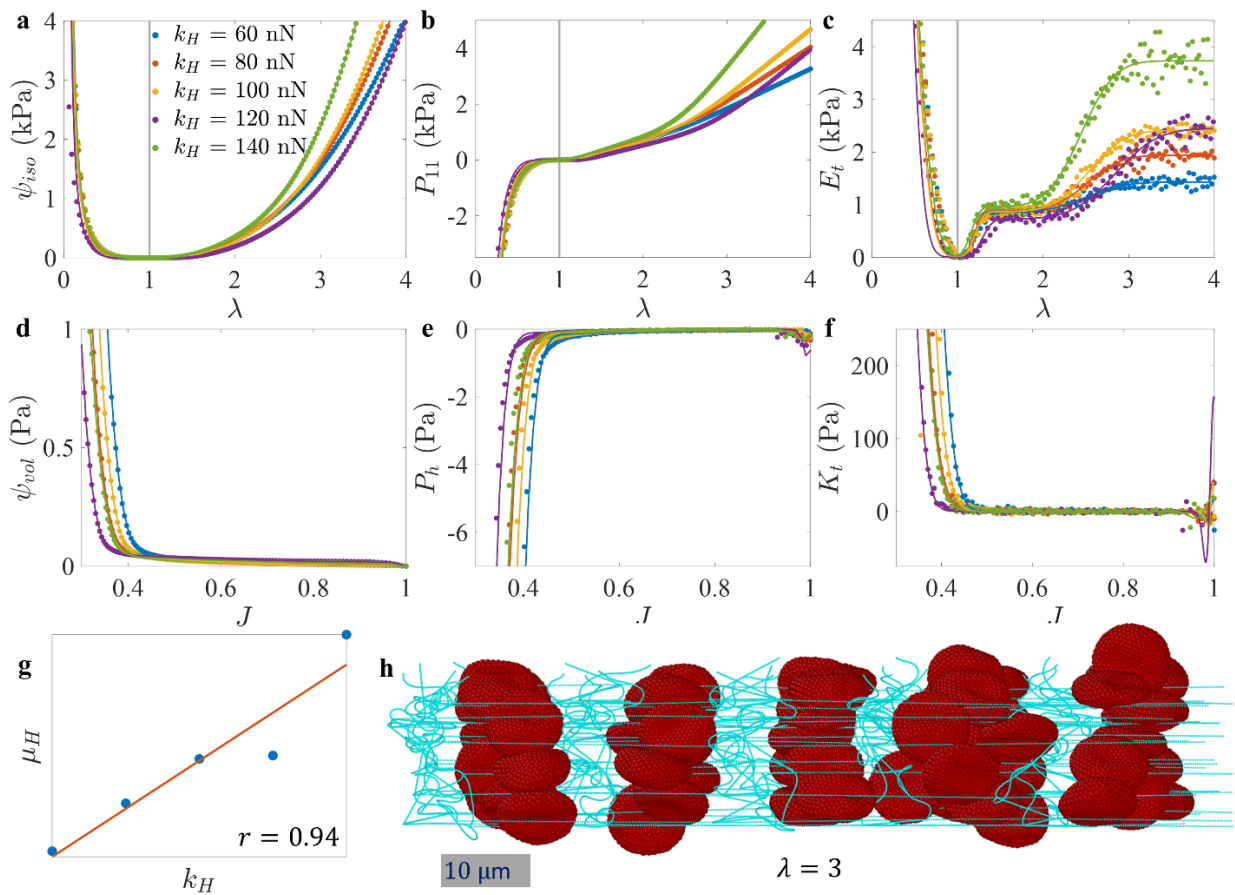

Figure 10: **Effect of fibrin fiber enthalpic stiffness**. **a-c** Strain-energy density, stress, and tangent modulus data for microscopic blood clot models having different fibrin fiber enthalpic stiffnesses in isochoric tension/compression. The strain-energy data were regressed by Equation 10 in our proposed model and one notable trend was identified. **d-f** Strain-energy density, stress, and tangent modulus data from microscopic blood clot models having different fibrin fiber enthalpic stiffnesses in volumetric contraction. The strain-energy data were regressed by Equation 10 in our proposed model and no notable trends were observed. **g** The trend found in a-c is plotted showing a correlation between fibrin fiber enthalpic stiffness and the hyperelastic blood clot stiffness term introduced to capture fibrin fiber enthalpic stiffness. **h** A pictures of a microscopic blood clot model is pictured at $\lambda = 3$ where the fibrin fiber enthalpic stiffness determines the clot stiffness.

The strain-energy density $\psi_{vol}$, hydrostatic stress $P_h$, and tangent volume modulus $K_t$ of the blood clot models subjected to volumetric contraction are shown in Figure 10d-f. The hyperelastic potential given by our proposed model fit these strain-energy density curves with $R^2$ values between 0.9978 and 0.9992. The fibrin fiber enthalpic stiffness, $k_H$, has no discernable effect on the volumetric strain-energy density, hydrostatic stress, and tangent volume modulus when $J > 0.6$, but the effect on those three values is distinguishable when $J < 0.6$. Compression caused the blood clot to extend along the normal plane of the compression direction. After that, the expansion yield stretches on fibers in normal direction, which will transit from entropic stretch into enthalpic stretch. Diverging from the entropic stretch, enthalpic stretch resembles a final status during the stretching process, wherein as strain increases, all fibers ultimately reach enthalpic stretch. Therefore, as the stretch continues, an increasing number of fibers transition to enthalpic stretch, which amplifies the effect of entropy stretch stiffness, leading to an "accumulated" effect. Therefore, as stretch persists, the influence of enthalpic stiffness gradually becomes apparent and eventually clearly distinguishable.

Figure 10g shows the correlation that was found between the fibrin fiber enthalpic stiffness, $k_H$, and the stiffness constants of Equation 10. The stiffness constant $\mu_H$ from Equation 5, describing the strain-energy contribution due to fibrin fiber enthalpic stiffness, was found to be directly proportional to $k_H$, with a Pearson correlation coefficient of 0.94. Figure 10h depicts a blood clot at a stretch where the enthalpic stiffness is relevant.

*5.6 Effect of fibrin fiber bending stiffness $B_s$*

The fibrin fiber bending stiffness, $B_s$, was varied from $4.28 \times 10^{-27}$ Nm$^2$ to $4.28 \times 10^{-23}$ Nm$^2$, these values correspond to fiber persistence lengths of 1 μm to 10 mm at body temperature (~310 K). The remaining adjustable parameters were held constant at $\phi_c \approx 0.56$, $\xi \approx 6.77$, $\tau = 1.15$ nN, $k_S = 29$ nN, $k_H = 90$ nN, $m = 0.85$, and $s = 0.11$. The strain-energy density $\psi_{iso}$, tensile/compressive stress $P_{11}$, and tangent elastic modulus $E_t$ of the blood clot models subjected to isochoric tension and compression are shown in Figure 11a-c. The hyperelastic potential given by our proposed model fit these strain-energy density curves with $R^2$ values between 0.9958 and 0.9999. The effect of fibrin fiber bending stiffness, $B_s$, on the isochoric strain energy density, $\psi_{iso}$, is not distinguishable in Figure 11a-c.

The strain-energy density $\psi_{vol}$, hydrostatic stress $P_h$, and tangent volume modulus $K_t$ of the blood clot models subjected to volumetric contraction are shown in Figure 11d-f. The hyperelastic potential given by our model fit these strain-energy density curves with $R^2$ values between 0.9988 and 0.9993. The effect of fibrin fiber bending stiffness, $B_s$, on the volumetric strain energy $\psi_{vol}$ is not distinguishable in Figure 11d-f. As the compression continues, the load is added to jammed RBCs, fiber buckling cannot have significant effect. Also, due to the huge difference in magnitude, bending energy does not exert a dominant effect.

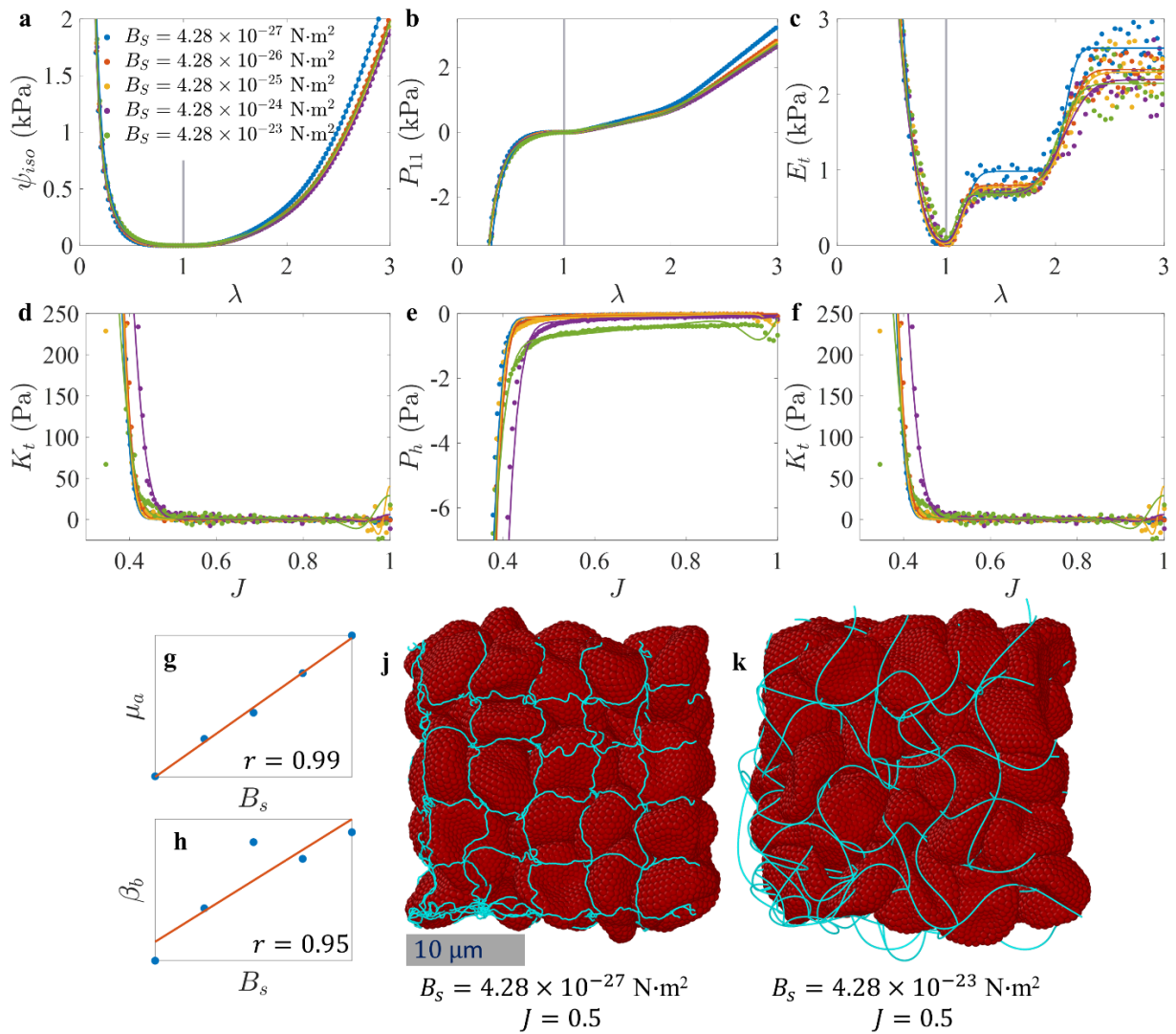

Figure 11: **Effect of fibrin fiber bending stiffness**. **a-c** Strain-energy density, stress, and tangent modulus data for microscopic blood clot models having different fibrin fiber bending stiffnesses in isochoric tension/compression. The strain-energy data were regressed by Equation 10 in our proposed model and notable trends were identified. **d-f** Strain-energy density, stress, and tangent modulus data from microscopic blood clot models having different fibrin fiber

bending stiffnesses in volumetric contraction. The strain-energy data were regressed by Equation 10 in our proposed model and notable trends were observed. **g-h** The trends found in **a-f** are plotted showing a correlation between fibrin fiber bending stiffness and the hyperelastic blood clot stiffness terms introduced to capture fiber alignment and buckling. **j-k** Pictures of microscopic blood clot models with different fibrin fiber bending stiffnesses are shown at the same volumetric contraction highlighting the effect of bending stiffness on fibrin mesh behavior.

Figure 11g-h shows the correlations between the fibrin fiber bending stiffness, $B_s$, and the stiffness constants of our proposed model. The stiffness constants $\mu_a$ and $\beta_b$ from Equations 3 and 6, describing the strain-energy contribution due to fibrin fiber alignment and fibrin fiber buckling, respectively, were found to be directly proportional to $B_s$, with Pearson correlation coefficients of 0.99 and 0.95. Figure 11j-k depicts volumetrically contracted blood clots with different fibrin fiber bending stiffnesses. The highly flexible fibrin fibers appear to be thermally slacked and the stiffer fibrin fibers are more reminiscent of buckled wires.

*5.7 Mechanistic formulation of strain energy density $\psi$ with microscopic inputs*

From the single variate analysis, we are able to see that the stiffness constants of Equation 10 in our proposed model had clear relationship with the microscopic properties of the simulated blood clot models. To complete the analysis, the correlations found in the previous sections and the data sets for each single variate analysis are combined and subjected to multiple linear regression. Subsequently, this series of multiple linear regressions, the following relationships were found for the stiffness and strain constants of Equations 3 through 5 (Figure 12).

Single analysis exhibits both merits and limitations. It surpasses multivariate analysis in discerning the presence of correlations and elucidating relationships between two variables. Nonetheless, the univariate approach falters when tasked with unveiling the comprehensive equation for a dependent variable influenced by multiple factors. Conversely, multivariate analysis encounters challenges in detecting correlations between two variables and delineating their precise relationship. Despite the complexity inherent in multivariate analysis, its potency shines through in assessing whether the impact of multiple variables is additive, multiplicative, intricately intertwined, or marked by multicollinearity. The amalgamation of these two approaches strategically addresses and compensates for their respective limitations.

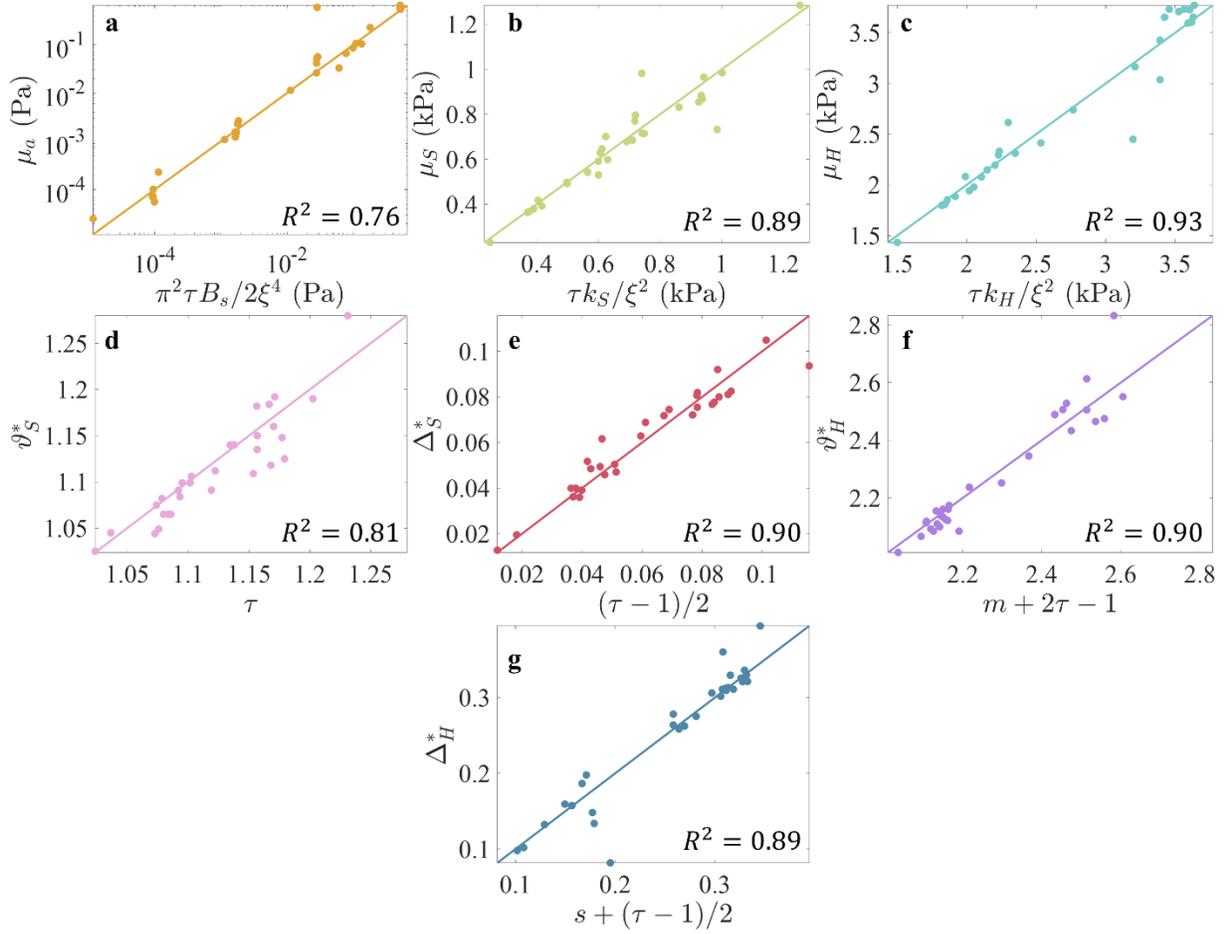

Figure 12: **Quantitative analysis of mechanistic equations for stiffness/strain parameters regarding fibrin fibers in tension. a-g** Mechanistic descriptions of the stiffness/strain parameters in Equations 3-5 found from multivariate analysis. The results of multivariate analysis provide reasonable confidence for the descriptions of the contribution of entropic/enthalpic stiffness of fibrin fibers to blood clot deformation given by **b-c** and the strain parameters given in **d-g**.

The shear modulus due to fiber alignment (Equation 3) was best described by an equation resembling the post-buckling stiffness of beams [51] (Figure 12a),

$$\mu_a = \frac{\pi^2 \tau B_s}{2\xi^4} \tag{19}$$

in which it is better understood as the stiffness of a bent fiber being straightened. The shear modulus due to fibrin fiber entropic and enthalpic stiffness was best described by the fiber stiffness, mesh size, and mesh tortuosity [15, 52] (Figure 12b-c),

$$\mu_S = \frac{\tau k_S}{\xi^2} \tag{20}$$

$$\mu_H = \frac{\tau k_H}{\xi^2} \tag{21}$$

this relationship is usually reported as $\mu = \frac{k}{\xi^2}$, where $\mu$ is the shear modulus of the mesh and $k$ is the stiffness of the fiber. Here, we found that the shear moduli also scale linearly with the mesh tortuosity. Mesh tortuosity typically exhibits a value highly close to 1, so this can easily be missed in the analysis of experimental or simulated stress-strain data. The mesh tortuosity not only effectively decreases the cross-sectional area of the clot, resulting in the expectation that $\mu \sim \tau^2$, but it also increases the disparity between macroscopic and microscopic strains, leading to the expectation that $\mu \sim \frac{1}{\tau}$. The combined effect is a linear dependence on $\tau$.

The strain parameters in Equations 4 and 5, describing the strain-energy dependence of blood clots on the entropic and enthalpic stiffness of fibrin fibers, were also elucidated from the multivariate analysis (Figure 12d-g),

$$\vartheta_S^* \approx \tau \tag{22}$$
$$\Delta_S^* \approx (\tau - 1)/2 \tag{23}$$
$$\vartheta_H^* \approx m + 2\tau - 1 \tag{24}$$
$$\Delta_H^* \approx s + (\tau - 1)/2 \tag{25}$$

without the insight gleaned from single variate analysis or a much larger data set, we report these relationships with lower confidence as approximations.

The following relationships were found for the stiffness and strain constants in Equations 6, 7, and 9 (Figure 13).

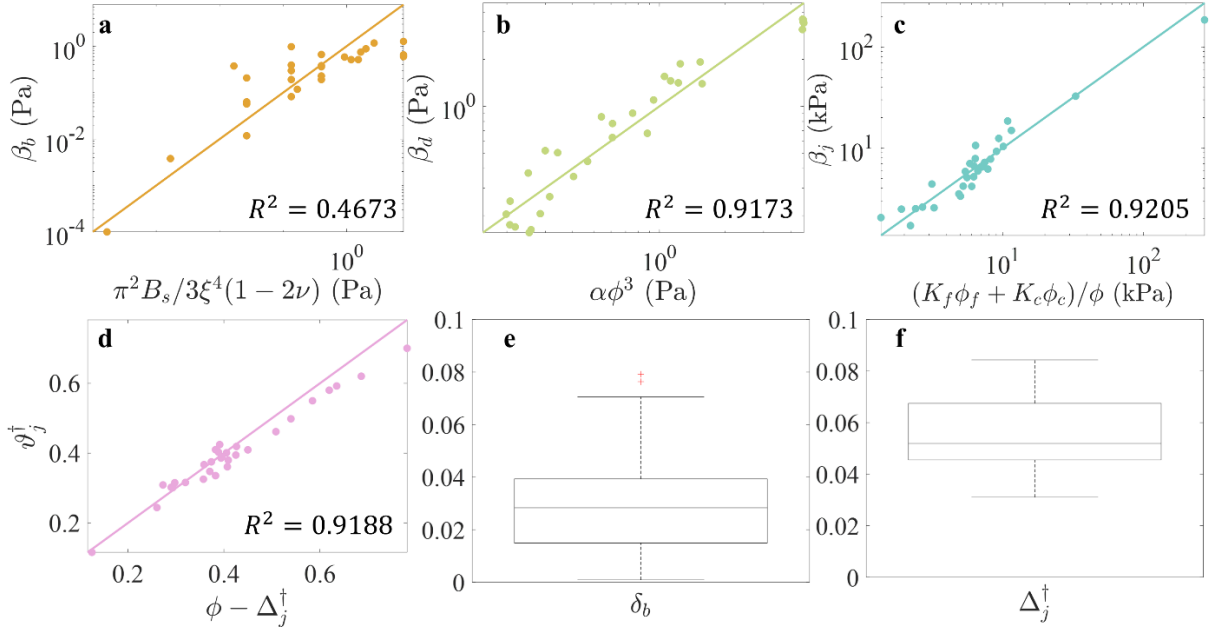

Figure 13: **Quantitative analysis of mechanistic equations for stiffness/strain parameters regarding blood clot volume change**. **A-d** Mechanistic descriptions of the stiffness/strain parameters in Equations 6, 7, and 9 found from multivariate analysis. The accuracy of the mechanistic equation found to describe the stiffness constant $\beta_b$ is too low report with confidence. **E-f** No trends were identified for the strain parameters $\delta_b$ and $\Delta_j^\dagger$.

The volumetric stiffness due to axial fiber compression and buckling was found to be,

$$B_b \approx \frac{\pi^2 B_s}{3\xi^4(1-2\nu)} \tag{26}$$

where $\nu$ is the Poisson's ratio and was introduced phenomenologically based on the theory of linearly elasticity (Figure 13a). This relationship is reported as an approximation due to the unacceptably low $R^2$ value. The poor description of $\beta_b$ could be attributed to one or more of several issues including: 1) a lack of information to describe $\beta_b$, some microscopic variables that we failed to consider when characterizing our models, 2) the buckling model given by Equation 6 could be a poor continuum description of fiber buckling, therefore, fitting it would not result in meaningful information, or 3) the simulations do not accurately capture the buckling behavior of fiber meshes. In any case, strain-energy term for fiber buckling given in Equation 6 and the mechanistic description given in Equation 26 should be held in suspicion.

The stiffness constant $\beta_d$ of the strain-energy term proposed to describe blood clot densification was found to be described by a power-law of the combined RBC and fibrin fiber volume fractions (Figure 13b),

$$\beta_d = \alpha \phi^3 \tag{27}$$

The stiffness constant $\beta_j$ of the strain-energy term proposed to describe blood clot jamming was found to resemble the rule of mixtures [53-55] (Figure 13c),

$$\beta_j = \frac{K_f \phi_f + K_c \phi_c}{\phi} \tag{28}$$

where $K_f$ and $K_c$ were introduced phenomenologically as the bulk modulus of the fibrin fibers and the bulk modulus of the RBCs, respectively. The strain parameter that marks the midpoint of the transition from blood clot densification to blood clot jamming, $\vartheta_j^\dagger$, from Equation 9 was found to be,

$$\vartheta_j^\dagger \approx \phi - \Delta_j^\dagger \tag{29}$$

However, no relationship was found for the strain parameters $\delta_b$ and $\Delta_j^\dagger$.

In conclusion, the single variate and multivariate analyses were able to provide some justification for the strain-energy terms $\psi_{align}$, $\psi_{entropic}$, $\psi_{enthalpic}$, $\psi_{densification}$, and $\psi_{jammed}$. Furthermore, the multivariate analysis was able to identify meaningful descriptions for the majority of the stiffness and strain constants in the aforementioned strain-energy terms. The multivariate analysis also warrants further investigation into the strain-energy term $\psi_{buckle}$. However, it did not offer support for this term, indicating that our proposed strain-energy function for fiber buckling might be incorrect.

## 6. Conclusions

This study introduced a hyperelastic potential for blood clots based on a novel fibrin fiber force-extension equation. Utilizing simple harmonic oscillators, one-sided harmonic potentials, and a Gaussian potential, the model are developed based on five key assumptions, including isotropic fiber distribution and non-inclusion of post-buckling potentials. The developed model, continuous in $C^0$, $C^1$, and $C^2$ forms with 13 parameters, showed superior performance in tension and shear analyses compared to the Fereidoonnezhad-McGarry model, though it was less effective in describing compression. Subsequent blood clot simulations helped identify mechanistic relationships for most of the stiffness/strain parameters, and highlighted a potential flaw in the fiber buckling term. This suggests a need for revising this aspect of the model. Comparative analysis with the Fereidoonnezhad-McGarry and Purohit et al. models revealed advantages and limitations of each. Our model, along with Purohit et al.'s, provides continuous derivatives and treats blood clots isotropically, in contrast to the piecewise, anisotropic approach of the Fereidoonnezhad-McGarry model. While our model and Purohit et al.'s offer mechanistic insights, the latter inaccurately describes large strain behaviors. Additionally, our model's stress and stiffness derivations are more complex than those of the other models but results in a smaller parameter space and system of equations than the Fereidoonnezhad-McGarry model.

Despite the recent advance of the surgery technique and the developments of anti-thrombotic treatments, the mortality and morbidity induced by the formation of undesired blood clots remain high. In this study, we introduce a novel hyperelastic model for blood clots that can advance our quantitative understanding of the relationship between the microscopic features of blood clots and their macroscopic behavior under deformation, thereby offering the potential to reveal the complex microscopic characteristics of clots during biomechanical analysis. This model holds the promise of establishing a new approach to elucidate the complex connections between microscopic observations and macroscopic experimental measurements. Overall, this research advances our understanding of blood clot mechanics, highlighting areas for further refinement.

**Conflict of Interest**

The authors declare that the research was conducted in the absence of any commercial or financial relationships that could be construed as a potential conflict of interest.

**Data Availability Statement**

# Appendix

*Appendix A: Right-sided and left-sided harmonic potential*

The strain-energy terms in Equations 4 and 5, describing the strain-energy due to entropic and enthalpic stretching of fibrin fibers, were defined by a right-sided harmonic potential, $\psi_{RSH} = \frac{\mu}{2} f^*\{\lambda; \vartheta^*, \Delta^*\} = \frac{\mu}{2} \int_1^\lambda \int_1^y \left(1 + \mathrm{erf}\left[\frac{x-\vartheta^*}{\Delta^*\sqrt{2}}\right]\right) dx dy$. The explicit equation for this right-sided harmonic potential is

$$\psi_{RSH} = \frac{\lambda^2}{2} - \lambda + \frac{1}{2} + \left(\frac{\vartheta^* \Delta^*}{\sqrt{2\pi}} + \frac{\Delta^*}{\sqrt{2\pi}} - \lambda \Delta^* \sqrt{\frac{2}{\pi}}\right) e^{-\left(\frac{\vartheta^*-1}{\sqrt{2}\Delta^*}\right)^2} + \left(\frac{\lambda \Delta^*}{\sqrt{2\pi}} - \frac{\vartheta^* \Delta^*}{\sqrt{2\pi}}\right) e^{-\left(\frac{\lambda-\vartheta^*}{\sqrt{2}\Delta^*}\right)^2} + \frac{1}{2}(2\lambda(\vartheta^* - 1) - (\vartheta^*)^2 + (\Delta^*)^2 + 1)\,\mathrm{erf}\left[\frac{1-\vartheta^*}{\sqrt{2}\Delta^*}\right] + \frac{1}{2}(\lambda^2 - 2\lambda\vartheta^* + (\vartheta^*)^2 + (\Delta^*)^2)\,\mathrm{erf}\left[\frac{\lambda-\vartheta^*}{\sqrt{2}\Delta^*}\right] + (\Delta^*)^2 \mathrm{erf}\left[\frac{\vartheta^*-1}{\sqrt{2}\Delta^*}\right] \tag{A1}$$

In general, the first derivative of $\psi_{RSH}$ with respect to $\lambda$ is

$$\frac{\partial \psi_{RSH}}{\partial \lambda} = \lambda + \Delta^* \sqrt{\frac{2}{\pi}} \left(e^{-\left(\frac{\lambda-\vartheta^*}{\sqrt{2}\Delta^*}\right)^2} - e^{-\left(\frac{\vartheta^*-1}{\sqrt{2}\Delta^*}\right)^2}\right) + (\vartheta^* - 1)\,\mathrm{erf}\left[\frac{1-\vartheta^*}{\sqrt{2}\Delta^*}\right] + (\lambda - \vartheta^*)\,\mathrm{erf}\left[\frac{\lambda-\vartheta^*}{\sqrt{2}\Delta^*}\right] - 1 \tag{A2}$$

The strain-energy term in Equation 9 describing the strain-energy due to blood clot jamming was defined by a left-sided harmonic potential, $\psi_{LSH} = \frac{\beta}{2} f^\dagger\{J; \vartheta^\dagger, \Delta^{*\dagger}\} = \frac{\beta}{2} \int_1^J \int_1^y \left(1 - \mathrm{erf}\left[\frac{x-\vartheta^\dagger}{\Delta^\dagger\sqrt{2}}\right]\right) dx dy$. The explicit equation for this left-sided harmonic potential is,

$$\psi_{LSH} = \frac{J^2}{2} - J + \frac{1}{2} + \left(J\Delta^\dagger \sqrt{\frac{2}{\pi}} - \frac{\vartheta^\dagger \Delta^\dagger}{\sqrt{2\pi}} - \frac{\Delta^\dagger}{\sqrt{2\pi}}\right) e^{-\left(\frac{\vartheta^\dagger-1}{\sqrt{2}\Delta^\dagger}\right)^2} + \left(\frac{\vartheta^\dagger \Delta^\dagger}{\sqrt{2\pi}} - \frac{J\Delta^\dagger}{\sqrt{2\pi}}\right) e^{-\left(\frac{J-\vartheta^\dagger}{\sqrt{2}\Delta^\dagger}\right)^2} - \frac{1}{2}(2J\vartheta^\dagger - (\vartheta^\dagger)^2 + (\Delta^\dagger)^2 - 1)\,\mathrm{erf}\left[\frac{1-\vartheta^\dagger}{\sqrt{2}\Delta^\dagger}\right] - \frac{1}{2}(J^2 - 2J\vartheta^\dagger + (\vartheta^\dagger)^2 + (\Delta^\dagger)^2)\,\mathrm{erf}\left[\frac{J-\vartheta^\dagger}{\sqrt{2}\Delta^\dagger}\right] - (J + (\Delta^\dagger)^2 - 1)\,\mathrm{erf}\left[\frac{\vartheta^\dagger-1}{\sqrt{2}\Delta^\dagger}\right] \tag{A3}$$

In general, the first derivative $\psi_{LSH}$ with respect to $\lambda_i$ is

$$\frac{\partial \psi_{LSH}}{\partial \lambda_i} = \lambda_q \lambda_r \left(\Delta^\dagger \sqrt{\frac{2}{\pi}} \left(e^{-\left(\frac{\vartheta^\dagger-1}{\sqrt{2}\Delta^\dagger}\right)^2} - e^{-\left(\frac{J-\vartheta^\dagger}{\sqrt{2}\Delta^\dagger}\right)^2}\right) + (1 - \vartheta^\dagger)\,\mathrm{erf}\left[\frac{1-\vartheta^\dagger}{\sqrt{2}\Delta^\dagger}\right] + (J - \vartheta^\dagger)\,\mathrm{erf}\left[\frac{J-\vartheta^\dagger}{\sqrt{2}\Delta^\dagger}\right] + \vartheta^\dagger - 2\right) \tag{A4}$$

where $q$ and $r$ are equal 2 and 3 if $i = 1$, 1 and 3 if $i = 2$, and 1 and 2 if $i = 3$.

*Appendix B: General formulations of stress*

The stresses in a hyperelastic material are found by considering the change in energy density due to deformation. Strain energy density, $\psi$, is a scalar quantity defined at every point in a material and its derivative with respect to various strain measures yields various stress measures. For a compressible hyperelastic material model, the principal 1st Piola Kirchhoff stresses (engineering stresses) is by definition, $P_i = \frac{\partial \psi}{\partial \lambda_i}$ with $i = 1, 2, 3$. For tension and compression, the principal directions align with the $x$, $y$, and $z$ axes, therefore $P_{11} = P_1$, $P_{22} = P_2$, and $P_{33} = P_3$.

The 1st Piola Kirchhoff stress may also be written as a spectral decomposition, $\boldsymbol{P} = \sum_{a=1}^{3} P_a \widehat{\boldsymbol{n}}_a \otimes \widehat{\boldsymbol{N}}_a$ where $P_a = \frac{\partial \psi}{\partial \lambda_a}$, $\lambda_a$ are the principal stretches, $\widehat{\boldsymbol{n}}_a$ are the normalized eigenvectors of $\boldsymbol{b}$, and $\widehat{\boldsymbol{N}}_a$ are the normalized eigenvectors of $\boldsymbol{C}$. This definition of 1st Piola Kirchhoff stress is convenient for deriving shear stress. For simple shear, the deformation gradient is expressed in matrix representation as, $\boldsymbol{F} = \begin{bmatrix} 1 & \gamma & 0 \\ 0 & 1 & 0 \\ 0 & 0 & 1 \end{bmatrix}$. The left Cauchy Green tensor for simple shear is, $\boldsymbol{b} = \boldsymbol{F}\boldsymbol{F}^T = \begin{bmatrix} \gamma^2 + 1 & \gamma & 0 \\ \gamma & 1 & 0 \\ 0 & 0 & 1 \end{bmatrix}$ and has the following normalized eigenvectors,

$$\widehat{n}_1 = \begin{bmatrix} \frac{\gamma+\sqrt{4+\gamma^2}}{2\sqrt{1+\frac{1}{4}\left(\gamma+\sqrt{4+\gamma^2}\right)^2}} \\ \frac{1}{\sqrt{1+\frac{1}{4}\left(\gamma+\sqrt{4+\gamma^2}\right)^2}} \\ 0 \end{bmatrix}, \widehat{n}_2 = \begin{bmatrix} \frac{\gamma-\sqrt{4+\gamma^2}}{2\sqrt{1+\frac{1}{4}\left(\gamma-\sqrt{4+\gamma^2}\right)^2}} \\ \frac{1}{\sqrt{1+\frac{1}{4}\left(\gamma-\sqrt{4+\gamma^2}\right)^2}} \\ 0 \end{bmatrix}, \widehat{n}_3 = \begin{bmatrix} 0 \\ 0 \\ 1 \end{bmatrix}$$

The right Cauchy Green tensor for simple shear is, $\boldsymbol{C} = \boldsymbol{F}^T \boldsymbol{F} = \begin{bmatrix} 1 & \gamma & 0 \\ \gamma & \gamma^2 + 1 & 0 \\ 0 & 0 & 1 \end{bmatrix}$ and has the following normalized eigenvectors,

$$\widehat{N}_1 = \begin{bmatrix} \frac{-\gamma+\sqrt{4+\gamma^2}}{2\sqrt{1+\frac{1}{4}\left(-\gamma+\sqrt{4+\gamma^2}\right)^2}} \\ \frac{1}{\sqrt{1+\frac{1}{4}\left(-\gamma+\sqrt{4+\gamma^2}\right)^2}} \\ 0 \end{bmatrix}, \widehat{N}_2 = \begin{bmatrix} \frac{\gamma-\sqrt{4+\gamma^2}}{2\sqrt{1+\frac{1}{4}\left(\gamma-\sqrt{4+\gamma^2}\right)^2}} \\ \frac{1}{\sqrt{1+\frac{1}{4}\left(\gamma-\sqrt{4+\gamma^2}\right)^2}} \\ 0 \end{bmatrix}, \widehat{N}_3 = \begin{bmatrix} 0 \\ 0 \\ 1 \end{bmatrix}$$

By definition $P_a = \frac{\partial \psi}{\partial \lambda_a}$ so the shear stress, $P_{12}$, is given by,

$$P_{12} = \left(\frac{\gamma+\sqrt{4+\gamma^2}}{2\sqrt{1+\frac{1}{4}\left(\gamma+\sqrt{4+\gamma^2}\right)^2}}\right)\left(\frac{1}{\sqrt{1+\frac{1}{4}\left(-\gamma+\sqrt{4+\gamma^2}\right)^2}}\right)\frac{\partial\psi}{\partial\lambda_1} + \left(\frac{\gamma-\sqrt{4+\gamma^2}}{2\sqrt{1+\frac{1}{4}\left(\gamma-\sqrt{4+\gamma^2}\right)^2}}\right)\left(\frac{1}{\sqrt{1+\frac{1}{4}\left(-\gamma-\sqrt{4+\gamma^2}\right)^2}}\right)\frac{\partial\psi}{\partial\lambda_2}$$

(A5)

Furthermore, the normal stress is given by

$$P_{11} = \frac{\left(-\gamma+\sqrt{4+\gamma^2}\right)\left(\gamma+\sqrt{4+\gamma^2}\right)}{4\sqrt{\left(1+\frac{1}{4}\left(\gamma+\sqrt{4+\gamma^2}\right)^2\right)\left(1+\frac{1}{4}\left(-\gamma+\sqrt{4+\gamma^2}\right)^2\right)}} P_1 + \frac{\left(-\gamma-\sqrt{4+\gamma^2}\right)\left(\gamma-\sqrt{4+\gamma^2}\right)}{4\sqrt{\left(1+\frac{1}{4}\left(\gamma-\sqrt{4+\gamma^2}\right)^2\right)\left(1+\frac{1}{4}\left(-\gamma-\sqrt{4+\gamma^2}\right)^2\right)}} P_2 \quad \text{(A6)}$$

The principal stretches are the square root of the eigenvalues of either $C$ or $b$ and were found to be $\lambda_1 = \sqrt{\frac{1}{2}\left(2+\gamma^2+\gamma\sqrt{4+\gamma^2}\right)}$, $\lambda_2 = \frac{1}{\lambda_1} = \sqrt{\frac{1}{2}\left(2+\gamma^2-\gamma\sqrt{4+\gamma^2}\right)}$, and $\lambda_3 = 1$. Upon careful examination and substitution, the equation for shear stress simplifies to $P_{12} = \frac{\lambda_1^2}{\lambda_1^2+1}\frac{\partial\psi}{\partial\lambda_1} - \frac{\lambda_2^2}{\lambda_2^2+1}\frac{\partial\psi}{\partial\lambda_2}$ and the equation for normal stress simplifies to $P_{11} = \sqrt{\frac{1}{\lambda_1^2+\lambda_2^2+2}}\left(\frac{\partial\psi}{\partial\lambda_1}+\frac{\partial\psi}{\partial\lambda_2}\right)$.

*Appendix C: 1st and 2nd principal engineering stresses for hyperelastic blood clot model*

To calculate the stress formulas from Appendix B, formulas for the first and second principal stresses are required. Recalling the definition $P_i = \frac{\partial\psi}{\partial\lambda_i}$ the principal stresses for the proposed hyperelastic blood clot model is given by the following series of equations,

$$P_{i,align} = \frac{\partial\psi_{align}}{\partial\lambda_i} = \mu_a(\lambda_i - 1) \tag{A7}$$

$$P_{i,entropic} = \frac{\partial\psi_{entropic}}{\partial\lambda_i} = \frac{1}{2}(\mu_S - \mu_a)\frac{\partial}{\partial\lambda_i}(f^*\{\lambda;\vartheta_S^*,\Delta_S^*\}) \tag{A8}$$

$$P_{i,enthalpic} = \frac{\partial\psi_{enthalpic}}{\partial\lambda_i} = \frac{1}{2}(\mu_H - \mu_S - \mu_a)\frac{\partial}{\partial\lambda_i}(f^*\{\lambda;\vartheta_H^*,\Delta_H^*\}) \tag{A9}$$

$$P_{i,buckle} = \frac{\partial\psi_{buckle}}{\partial\lambda_i} = \frac{\beta_b}{\delta_b}\lambda_q\lambda_r(J-1)e^{\frac{1}{2}-\left(\frac{J-1}{\sqrt{2}\delta_b}\right)^2} \tag{A10}$$

$$P_{i,densify} = \frac{\partial\psi_{densify}}{\partial\lambda_i} = \beta_d\lambda_q\lambda_r(J-1) \tag{A11}$$

$$P_{i,jamming} = \frac{\partial\psi_{jamming}}{\partial\lambda_i} = \frac{\beta_j}{2}\frac{\partial}{\partial\lambda_i}(f^\dagger\{J;\vartheta_j^\dagger,\Delta_j^{*\dagger}\}) \tag{A12}$$

where $q$ and $r$ are equal 2 and 3 if $i = 1$, 1 and 3 if $i = 2$, and 1 and 2 if $i = 3$. The derivatives $\frac{\partial}{\partial\lambda_i}(f^*\{\lambda;\vartheta_H^*,\Delta_H^*\})$ and $\frac{\partial}{\partial\lambda_i}(f^\dagger\{J;\vartheta_j^\dagger,\Delta_j^{*\dagger}\})$ are given by Equations A2 and A4. The total principal stress is given by,

$$P_i = P_{i,align} + P_{i,entropic} + P_{i,enthalpic} + P_{i,buckle} + P_{i,densify} + P_{i,jamming} \tag{A13}$$

which can be used to solve for the tensile and compressive stresses directly and incorporated in Equations A5 and A6 to calculate the shear and normal stress due to simple shear.